\documentclass[sigconf]{acmart}

\usepackage{xspace}
\usepackage{xcolor,colortbl}
\usepackage{colortbl}
\usepackage{pgfplotstable}
\usepackage[euler]{textgreek}
\usepackage{pgfplots}
\usepackage{svg}
\usepackage{subfig}
\usepackage{graphicx}
\pgfplotsset{compat=1.16}
\usepackage{tikz}
\usepackage{bchart}
\usepackage{xurl}
\usetikzlibrary{decorations.pathreplacing}

\newcommand{\embu}{\mathbf{p}}
\newcommand{\embi}{\mathbf{q}}
\newcommand{\sembu}{p}
\newcommand{\sembi}{q}

\definecolor{mygreen}{HTML}{2ca02c}
\definecolor{myred}{HTML}{db1b1b}
\definecolor{myblue}{HTML}{1f77b4}
\definecolor{myorange}{HTML}{ff7f0e}

\definecolor{MP}{HTML}{003f5c}
\definecolor{SL}{HTML}{2f4b7c}
\definecolor{iA}{HTML}{665191}
\definecolor{NM}{HTML}{a05195}
\definecolor{MF}{HTML}{d45087}
\definecolor{ES}{HTML}{f95d6a}
\definecolor{RP}{HTML}{ff7c43}
\definecolor{PS}{HTML}{ffa600}

\definecolor{MP2}{HTML}{5c0000}
\definecolor{SL2}{HTML}{740024}
\definecolor{iA2}{HTML}{871147}
\definecolor{NM2}{HTML}{922a6e}
\definecolor{MF2}{HTML}{934497}
\definecolor{ES2}{HTML}{8660be}
\definecolor{RP2}{HTML}{677ce2}
\definecolor{PS2}{HTML}{0097ff}

\def\movielens{\textit{MovieLens-1M}\xspace}
\def\pinterest{\textit{Pinterest}\xspace}
\def\easer{EASE\textsuperscript{R}\xspace}
\def\rp3b{RP\textsuperscript{3}\textbeta\xspace}

\usepackage{tabularx,booktabs}
\newcommand{\ra}[1]{\renewcommand{\arraystretch}{#1}}
\usepackage{multirow}

\AtBeginDocument{%
  \providecommand\BibTeX{{%
    \normalfont B\kern-0.5em{\scshape i\kern-0.25em b}\kern-0.8em\TeX}}}

\copyrightyear{2021}
\acmYear{2021}
\setcopyright{acmcopyright}\acmConference[RecSys '21]{Fifteenth ACM Conference
on Recommender Systems}{September 27-October 1, 2021}{Amsterdam, Netherlands}
\acmBooktitle{Fifteenth ACM Conference on Recommender Systems (RecSys '21),
September 27-October 1, 2021, Amsterdam, Netherlands}
\acmPrice{15.00}
\acmDOI{10.1145/3460231.3475944}
\acmISBN{978-1-4503-8458-2/21/09}

\begin{document}
\title[Reenvisioning Collaborative Filtering vs Matrix Factorization]{Reenvisioning the comparison between Neural Collaborative Filtering and Matrix Factorization}

\author{Vito Walter Anelli}
\authornote{Authors are listed in alphabetical order. Corresponding authors: Vito Walter Anelli (\url{vitowalter.anelli@poliba.it}) and Claudio Pomo (\url{claudio.pomo@poliba.it}).}
\email{vitowalter.anelli@poliba.it}
\affiliation{\institution{Politecnico di Bari, Italy}
    \country{}
 }

\author{Alejandro Bellogín}
\email{alejandro.bellogin@uam.es}
 \affiliation{
 \institution{Universidad Autónoma de Madrid, Spain}
 \country{}
 }

\author{Tommaso Di Noia}
\email{tommaso.dinoia@poliba.it}
\affiliation{\institution{Politecnico di Bari, Italy}
\country{}
 }

\author{Claudio Pomo}
\authornotemark[1]
\email{claudio.pomo@poliba.it}
\affiliation{\institution{Politecnico di Bari, Italy}
\country{}
 }


\renewcommand{\shortauthors}{Anelli et al.}

\begin{abstract}
Collaborative filtering models based on matrix factorization and learned similarities using Artificial Neural Networks (ANNs) have gained significant attention in recent years.
This is, in part, because ANNs have demonstrated very good results in a wide variety of recommendation tasks.
However, the introduction of ANNs within the recommendation ecosystem has been recently questioned, raising several comparisons in terms of efficiency and effectiveness.
One aspect most of these comparisons have in common is their focus on accuracy, neglecting other evaluation dimensions important for the recommendation, such as novelty, diversity, or accounting for biases.
In this work, we replicate experiments from three different papers that compare Neural Collaborative Filtering (NCF) and Matrix Factorization (MF), to extend the analysis to other evaluation dimensions.
First, our contribution shows that the experiments under analysis are entirely reproducible, 
and we extend the study including other accuracy metrics and two statistical hypothesis tests.
Second, we investigated the Diversity and Novelty of the recommendations, showing that MF provides a better accuracy also on the long tail, although NCF provides a better item coverage and more diversified recommendation lists.
Lastly, we discuss the bias effect generated by the tested methods. They show a relatively small bias, but other recommendation baselines, with competitive accuracy performance, consistently show to be less affected by this issue.
This is the first work, to the best of our knowledge, where several complementary evaluation dimensions have been explored for an array of state-of-the-art algorithms covering recent adaptations of ANNs and MF. 
Hence, we aim to show the potential these techniques may have on beyond-accuracy evaluation while analyzing the effect on reproducibility these complementary dimensions may spark.
The code to reproduce the experiments is publicly available on GitHub at \em{\url{https://github.com/sisinflab/Reenvisioning-the-comparison-between-Neural-Collaborative-Filtering-and-Matrix-Factorization}}.

\end{abstract}



\maketitle


\section{Introduction}
Artificial Neural Networks (ANNs) are ubiquitous in many research areas in recent years.
Recommender Systems (RS) is a paradigmatic example where these techniques have been applied, in part because they open up possibilities in domains where classical techniques have difficulties in understanding the item content --- i.e., video or image recommendation ---, but also because they allow to extract complex patterns that, in principle, are not captured by more simple methods~\cite{DBLP:journals/csur/ZhangYST19}.
However, recent research challenges how useful these techniques are in the context of RS, and where their advantages really lie~\cite{tois,rendle,DBLP:conf/cikm/DacremaPCJ20,DBLP:conf/recsys/JannachMO20}.

More specifically, these studies have compared ANNs techniques against classical personalized algorithms --- mostly matrix factorization or nearest neighbors ---, emphasizing the lack of well-tuned baselines or incorrect, incomplete, or even unfair experimental conditions evidenced in the literature.
Nonetheless, while these conclusions are useful to move forward on understanding when ANNs should be applied in recommendation, they neglect evaluation dimensions that are important in the RS community, such as diversity, novelty, coverage, and so on~\cite{DBLP:reference/sp/GunawardanaS15}, since most of the authors have focused, so far, on the precision/accuracy of the recommended items produced by those methods. 

In this context, we aim to bridge this gap and compare ANNs against classical RS under several evaluation dimensions.
With this goal in mind, we focus on a recent paper~\cite{rendle} where the authors showed how proper hyperparameter selection could make simple operations like a dot product outperform similarity learning through ANNs.
We have the following two main goals: 
first, \textbf{replicating} the aforementioned paper, since it has a salient characteristic where the authors used in their tables results from other papers (claiming they used comparable evaluation settings and tuning, something that too often is not true as it is difficult to do properly~\cite{DBLP:conf/recsys/SaidB14}); 
once we are able to replicate these results, we \textbf{reproduce} them under different situations.
In particular, we report beyond-accuracy evaluation metrics, to explore the extent these methods behave on complementary dimensions they have not been optimised for, or whose results have not been reported about.

Our main contributions are two-fold: on the one side, we corroborate the results reported recently in \cite{rendle} where ANNs are outperformed by simple modifications on classical algorithms; moreover, we complement these observations with additional experimental dimensions, showing more accuracy metrics and their corresponding statistical analysis, together with novelty, diversity, and bias measurements, which allow us to provide a more complete overview of the performance of these algorithms when compared against ANNs and, hence, a better understanding of when and how these approaches might be useful.

\section{Background and formulation}
In this section we formalize, first, the recommendation problem and later review Matrix Factorization and Neural Collaborative Filtering approaches.
The notation used herein is summarised as follows.
Matrices are denoted by uppercase letters $A$, vectors by lowercase bold letters $\mathbf{b}$, scalars by lowercase letters $a$.
We denote the concatenation of the vectors $\mathbf{b}$ and $\mathbf{c}$ by $[\mathbf{b}, \mathbf{c}]$.
Let there be a pool of users ($U$) to recommend to and a catalog of items ($I$) to recommend from. 
A recommendation algorithm returns a \textit{score} for a given user-item pair that corresponds to the estimated degree of \textit{satisfaction} for the user enjoying that item.
In this work we focus on a specific kind of recommendation algorithms, where two $d$-dimensional embedding vectors, $p$ and $q$, are combined into a single score. 
Conventionally, $p$ represents the embedding of a user, $q$ the embedding of an item, and $\phi(p, q)$ ($\phi : \mathbb{R}^d \times \mathbb{R}^d \to \mathbb{R}$) the similarity of the user to the item.

Matrix Factorization (MF) is a famous and classical example of model-based Collaborative Filtering methods~\cite{DBLP:reference/sp/KorenB15}. The algorithm learns a latent representation of items and users, whose linear interactions aim to explain the observed feedback. 
There are several variations of MF proposed in the literature, and a comprehensive review would deserve a specific study that is out of the scope of this work.
However, to provide the reader an intuition of how much the factorization strategy has been disruptive in recent years, we briefly review the works that are, in our humble opinion, the most representative or the ones that can show the myriad of possible applications of factorization models.

The first examples of factorization models were soon recognized as state-of-the-art models. Among these pioneering works, there could be found SVD~\cite{DBLP:reference/sp/KorenB15}, PureSVD~\cite{DBLP:conf/recsys/CremonesiKT10}, SVD++~\cite{DBLP:conf/kdd/Koren08}, PMF~\cite{DBLP:conf/nips/SalakhutdinovM07,DBLP:conf/icml/SalakhutdinovM08a}, NNMF~\cite{DBLP:journals/tii/LuoZXZ14}, and SLIM~\cite{DBLP:conf/icdm/NingK11}.
Among the several methods on matrix factorization, Rendle's work has heavily influenced the evolution of the factorization models.
In detail, BPR-MF~\cite{DBLP:conf/uai/RendleFGS09} deserves particular attention because it boosted the MF research, and it is still considered as a state-of-the-art model.
For completeness, Rendle also proposed Factorization Machines~\cite{DBLP:conf/icdm/Rendle10} that generalize the factorization approach.
The biggest criticism of MF approaches, however, lies in their linearity. To address this concern, a recently popularized trend in the community of recommender systems is using deep neural architectures with deep neural networks that can model the non-linearity in data through nonlinear activation functions. 
In this respect, Neural Collaborative Filtering~\cite{ncf} and Neural Factorization Machines~\cite{DBLP:conf/sigir/0001C17} have been recently proposed to overcome the inability of MF to capture non-linearities.
Furthermore, Attentional Factorization Machines~\cite{DBLP:conf/ijcai/XiaoY0ZWC17} use an attention network to learn the importance of feature interactions. 
Factorization models have been specialized for a variety of tasks such as Active-Learning~\cite{DBLP:journals/tkde/ZhuLHWGLC20}, Context-aware~\cite{DBLP:conf/recsys/JuanZCL16}, Cross-domain~\cite{DBLP:journals/umuai/Fernandez-Tobias19}, Knowledge-aware~\cite{DBLP:conf/semweb/AnelliNSRT19,AnelliMinorTKDE}, and even explainable~\cite{DBLP:conf/sigir/ZhangL0ZLM14} recommendation.

In particular, Neural Collaborative Filtering~\cite{ncf} is one of the most representative recommendation approaches, which aims to estimate unknown user-item preference scores by exploiting deep neural networks~\cite{DBLP:journals/csur/ZhangYST19}. 
Since Artificial Neural Networks (ANNs) can approximate any continuous function on a compact set as long as the ANN has enough hidden states~\cite{DBLP:journals/mcss/Cybenko89},  
\citet{ncf} propose to exploit ANNs to learn the affinity between $p$ and $q$. 
Let $\Phi(\cdot)$ be the transformation function of the deep neural network defined as $\Phi: \mathbb{R}^{\text{dim}(p) + \text{dim}(q)} \to \mathbb{R}^d$, \citeauthor{ncf} propose to concatenate the two embeddings and predict the score as follows:
\begin{align}
    \psi^{\text{MLP}}(\embu, \embi) := \Phi([\embu, \embi]).
\end{align}
Additionally, \citeauthor{ncf} defines a \textit{generalized} matrix factorization model, in which $p$ and $q$ are combined using element-wise multiplication ($\odot$):
\begin{align}
    \psi^{\text{GMF}}(\embu, \embi) := \embu \odot \embi,
\end{align}
Finally,~\citeauthor{ncf} propose a comprehensive model, named NeuMF, that combines the two previous approaches together:
\begin{align}
\label{eq:neumf}
    \psi^{\text{NeuMF}}(\embu, \embi) := \psi^{\text{MLP}}(\embu, \embi) + \psi^{\text{GMF}}(\embu^{\prime}, \embi^{\prime}),
\end{align}
where the prime symbol ($\phantom{ }^{\prime}$) suggests that those embeddings might have a different size and are, in fact, different from the former ones.
Finally, a careful reader may have noticed that $\psi$ has an output dimension of $d$.
This is correct, since~\citeauthor{ncf} applies a final prediction layer on top of them:
\begin{align}
    \phi^{\text{NCF}} := \sigma(\mathbf{W} \cdot \psi(\embu, \embi))
\end{align}
where $\sigma$ is an activation function, and $W$ is an additional weight matrix that is learned along with the other model parameters.

More recently, \citet{rendle} define the embeddings as model parameters, and the affinity between $p$ and $q$ is modeled by means of a dot product:
\begin{align}
    \phi^{\text{dot}}(\embu, \embi) := b_g + b_p + b_q + \sum_{f=1}^d \sembu_f \sembi_f ,
\end{align}
where $g_b$, $b_p$, and $b_q$ denote the global, user, and item bias, respectively.
In this way, \cite{rendle} presents a direct comparison between ANNs and MF by changing the underlying operation between the embeddings, while keeping everything else comparable.

\section{Replication of prior experiments: settings and results}
This section focuses on describing how the replication of the experiments from papers~\citet{rendle,ncf,tois} has been set up. 
It starts by defining the evaluation protocol applied to compare Neural Collaborative Filtering (NCF) and Matrix Factorization (MF) against the baselines in their respective works.


\subsection{Settings}\label{sec:settings}
Although this study involves the replication of the results from three different studies, this paper mainly aims to replicate the results from~\citet{rendle}.
In~\citeauthor{rendle}, the authors retrieve the already split datasets from the original NCF repository\footnote{\url{https://github.com/hexiangnan/neural_collaborative_filtering}}.
Specifically,~\citet{ncf} provide a split version of \movielens and \pinterest. 
To split these well-known datasets, the authors adopt a temporal leave-one-out policy, moving the last user interaction into the test set.
Furthermore, they binarize \movielens to make the two datasets coherent with implicit feedback.
Finally, they evaluate the methods on a shortlist of $101$ candidate items for each user.
This list comprises one relevant item (i.e., the transaction in the test set) and $100$ negative items randomly sampled from not consumed items.
In~\citet{ncf} and~\citet{rendle}, the authors evaluate the performance on top-$10$ recommendation lists computing Hit-Rate (HR) and Normalized Discounted Cumulative Gain (nDCG).
The first estimates how many users have the withheld item in the top-$10$. 
The second measures the capability of the methods to rank the relevant item. In the following, the formulation of nDCG as presented in~\citet{DBLP:conf/kdd/KricheneR20} is adopted since it is the same one adopted in~\citet{rendle}.
Moreover,~\citet{ncf} and~\citet{rendle} select the best models, for each recommendation system, according to HR@10. In this paper, the model selection follows the same strategy.

The present study involved the implementation of seven recommendation methods.
MF implementation was designed accordingly to~\citet{rendle} (also provided as a public repository\footnote{\url{https://github.com/google-research/google-research/tree/master/dot_vs_learned_similarity}}).
Regarding NeuMF, the implementation refers to~\citet{ncf}.
Finally, for the five remaining baselines, the implementation refers to~\citet{tois} since it is the source for some of the results reported in~\citet{rendle}.
More specifically, the five implemented baselines are Slim~\cite{DBLP:conf/icdm/NingK11}, iALS~\cite{DBLP:conf/icdm/HuKV08}, PureSVD~\cite{DBLP:conf/recsys/CremonesiKT10}, \easer~\cite{DBLP:conf/www/Steck19}, and \rp3b~\cite{DBLP:journals/tiis/PaudelCNB17}. 
According to the investigation provided by the authors~\cite{tois}, we replicate the baseline training exploiting the best hyperparameters found in the additional material\footnote{\url{https://github.com/MaurizioFD/RecSys2019_DeepLearning_Evaluation/blob/master/DL_Evaluation_TOIS_Additional_material.pdf}}.

To summarize, this paper replicates seven different recommendation algorithms from three different works: \cite{rendle}, \cite{ncf}, and \cite{tois}.
The Elliot recommendation framework~\cite{DBLP:journals/corr/abs-2103-02590} is adopted as the benchmarking framework. 
Elliot provides an out-of-the-box recommendation pipeline.
The tested models have been implemented as external models to grant complete adherence to the original implementations. 
All the implemented models and configuration files are publicly available\footnote{\url{https://github.com/sisinflab/Reenvisioning-the-comparison-between-Neural-Collaborative-Filtering-and-Matrix-Factorization}} to provide a complete reproducibility environment with an ad-hoc version of Elliot.

\subsection{Results}
The first set of experiments aims to replicate Table 1 from~\citet{rendle}.
In that table, the authors compare Neural Matrix Factorization (NeuMF) and MF with a shortlist of baselines: Popularity, SLIM, and iALS.
In detail, the authors report from~\citet{tois} the results for Popularity, SLIM, NeuMF, and iALS.
Instead, MF is trained using the publicly available implementation they provide.
Overall, this table questions the prominence of NeuMF and shows the high performance achieved by MF.

Hence, Table~\ref{tab:basic_res} replicates and extends the results provided in Table 1 from~\citet{rendle}.
In this study, all the recommendation algorithms have been retrained according to the best hyperparameters provided in~\citet{tois}. 
Specifically, Table~\ref{tab:basic_res} reports HR and nDCG values for \movielens and \pinterest datasets, respectively.
The careful reader may have noticed that, for each dataset, both replicated and original results are reported.
Original results columns are marked with references to the source papers.
\citet{tois} also consider other recommendation algorithms. Interested in a more comprehensive comparison, we have selected \easer, \rp3b, and PureSVD for further replication.
Finally, since~\citet{tois} do not consider Matrix Factorization, no confusion arises regarding the origin of the results.
Interestingly, Table~\ref{tab:basic_res} further confirms the findings of the original experiments showing that MF consistently overcomes the other baselines.
It is worth mentioning how well the new experiments approximate the original ones.


\begingroup
\setlength{\tabcolsep}{1pt}
\begin{table*}
\centering
\caption{Comparison of NeuMF and MF with various baselines with cutoff @10. The table replicates (and compare with) the results from~\citet{rendle,tois}. The best results are highlighted in bold, the
second best results is underlined. The columns with the $\Delta$ symbol indicate the absolute variation (for each metric) between the values of the experiments reproduced and those reported in the articles by~\citet{tois} and ~\citet{rendle}.}\label{tab:basic_res}

\resizebox{\linewidth}{!}{\begin{tabular}{l@{\hskip 10pt}cc@{\hskip 6pt}cc@{\hskip 6pt}rr@{\hskip 10pt}|@{\hskip 6pt}cc@{\hskip 6pt}cc@{\hskip 6pt}rr}

\toprule

\multirow{2}{*}{Method} & \multicolumn{2}{c@{\hskip 6pt}}{\movielens} & \multicolumn{2}{c@{\hskip 6pt}}{\movielens~\cite{rendle,tois}} & \multicolumn{2}{c@{\hskip 6pt}|@{\hskip 6pt}}{$\Delta$  \movielens}& \multicolumn{2}{c@{\hskip 6pt}}{\pinterest} & \multicolumn{2}{c}{\pinterest~\cite{rendle,tois}}& \multicolumn{2}{c}{$\Delta$  \pinterest}\\

\cmidrule(r{6pt}){2-3} \cmidrule(r{6pt}){4-5} \cmidrule(r{12pt}){6-7} \cmidrule(r{6pt}){8-9} \cmidrule{10-11} \cmidrule(l{6pt}){12-13}
& \multicolumn{1}{c@{\hskip 6pt}}{nDCG} & \multicolumn{1}{c@{\hskip 6pt}}{HR} & \multicolumn{1}{@{\hskip 6pt}c@{\hskip 6pt}}{nDCG} & \multicolumn{1}{c@{\hskip 6pt}}{HR} & \multicolumn{1}{@{\hskip 6pt}c@{\hskip 6pt}}{nDCG} & \multicolumn{1}{c|@{\hskip 6pt}}{HR} & \multicolumn{1}{c@{\hskip 6pt}}{nDCG} & \multicolumn{1}{c@{\hskip 6pt}}{HR} & \multicolumn{1}{c@{\hskip 6pt}}{nDCG} & \multicolumn{1}{c@{\hskip 6pt}}{HR} & \multicolumn{1}{c@{\hskip 6pt}}{nDCG} & \multicolumn{1}{c@{\hskip 6pt}}{HR}\\ \midrule

MostPop & 0.2542& 0.4535 & 0.2543 & 0.4535 &$-1\cdot10^{-04}$ & 0 & 0.1410 & 0.2743& 0.1409& 0.2740& $1\cdot10^{-04}$ & $3\cdot10^{-04}$\\
SLIM & 0.4480& 0.7164& 0.4468 & 0.7162& $1.2\cdot10^{-03}$ & $2\cdot10^{-04}$ & 0.5615 &0.8696 &0.5601 & 0.8679& $1.4\cdot10^{-03}$ & $1.7\cdot10^{-03}$ \\
iALS & 0.4385& 0.7123 & 0.4383& 0.7111& $2\cdot10^{-04}$ & $1.2\cdot10^{-03}$ & 0.5587 & 0.8766& 0.5590& 0.8762& $3\cdot10^{-04}$ & $4\cdot10^{-04}$\\
NeuMF & 0.4211 & 0.6952 & 0.4349& 0.7093 &$-1.38\cdot10^{-02}$ & $-1.41\cdot10^{-02}$ & 0.5480 & 0.8704 & 0.5576& 0.8777& $-9.6\cdot10^{-03}$ & $-7.3\cdot10^{-03}$\\
MF & \textbf{0.4545}& \textbf{0.7310} & \textbf{0.4523}& \textbf{0.7294}& $2.2\cdot10^{-03}$ & $1.6\cdot10^{-03}$ & \textbf{0.5776}& \textbf{0.8898}& \textbf{0.5794} & \textbf{0.8895}& $1.8\cdot10^{-04}$ & $3\cdot10^{-04}$\\
\midrule
\easer & \underline{0.4494}& \underline{0.7192}& \underline{0.4494} & \underline{0.7192} & 0 & 0 & 0.5605& 0.8684 &0.5604 &0.8684& 0 & 0\\
\rp3b & 0.4011& 0.6758 & 0.4011 & 0.6758 & 0 & 0 & \underline{0.5685}& \underline{0.8796} & \underline{0.5685} & \underline{0.8796}& 0 & 0\\
PureSVD & 0.4299& 0.6926 & 0.4303 & 0.6937 & $-4\cdot10^{-04}$ & $-1.1\cdot10^{-03}$ & 0.5233& 0.8261 & 0.5241 & 0.8268& $-8\cdot10^{-04}$ & $-7\cdot10^{-04}$ \\
\bottomrule
\end{tabular}}
\end{table*}
Nonetheless, \citet{rendle} clearly state, in Table 1, that MF results are reported from (their) Figure 2.
That figure compares MF, Learned Similarity (MLP), NeuMF, and pretrained NeuMF, considering different embedding sizes.
However, the results reported from~\citeauthor{rendle} (except for MF) are from~\citet{ncf}.
Therefore, to conduct a thorough replication, we herein replicate some pivotal experiments reported in that figure.
In detail, we have decided to replicate six MF experiments (three embedding sizes for each dataset) and eight NeuMF experiments (four embedding sizes for each dataset).
For MF, we considered $32$, $128$, and $192$ as embedding sizes.
For NeuMF, we considered $24$, $48$, $96$, and $192$ as embedding sizes (according to Appendix $3$ from~\citet{rendle}).

Thus, Figure~\ref{fig:base_results} reports the original values from~\citet{ncf} regarding Learned Similarity (MLP) and pretrained NeuMF and reports our replicated experiments' results for MF and NeuMF.
It is noteworthy mentioning that our MF experiments overlap with~\citeauthor{rendle} showing that the MF curves dominate the others.
However, the NeuMF curve shows different behavior from~\citeauthor{ncf}
It is even more interesting to notice that the NeuMF experiment with $48$ as the embedding size is also reported by~\citet{tois}, and the results are very close to ours. 

Reviewing \pinterest results, MF confirms to be the best model in terms of HR and nDCG. 
Even in this context, NeuMF never overcomes the MF models. 
Actually, NeuMF reaches the best performance in terms of HR and nDCG for the NeuMF model with a number of factors equal to 16, according to the findings provided by~\citet{tois}. 
Nonetheless, increasing that number of factors, we witness a performance decrease: both HR and nDCG decrease as the number of factors increases. 
Furthermore, Table \ref{tab:basic_res} reports the overall results for the methods involved in the investigation. 
These outcomes confirm the evidence shown by~\citet{rendle}: also other MF-based methods, like SLIM and iALS, outperform NeuMF. 
Beyond MF, also \easer\ provides a very notable performance. 
PureSVD behaves similarly to NeuMF. 
Finally, \rp3b does not appear competitive as the other models in the investigation: its performance is consistently worse than the others.
All these findings further confirm the results provided in~\citet{tois}.
\begin{figure*}
\centering
\begin{tikzpicture}[scale=1,line width=1pt,every node/.style={outer sep=0}]
\begin{axis}[
    height=15em,
    width=0.47\textwidth,
    title={Movielens},
    legend style={nodes={scale=0.6, transform shape}},
    legend pos=south east,
    grid=major,
    xlabel=Embedding dimension,
    ylabel=HR@10,
    xmin=10, 
    xmax=256,
    xmode=log,
    log basis x={2},
    x tick label style={/pgf/number format/1000 sep=\,},
    log ticks with fixed point,
    ymin=0.550, ymax=0.750,  
    ytick={0.550,0.575,0.600,0.625,0.650,0.675,0.700,0.725,0.750},
    y tick label style={
        /pgf/number format/.cd,
        fixed,
        fixed zerofill,
        precision=3,
        /tikz/.cd
    }
    ]
\addplot[mark=*,myblue, line width=0.75pt, mark size=0.7] plot coordinates {
    (32,0.71821)
    (128,0.730629)
    (192,0.73096)
    
};
\addlegendentry{Matrix Factorization (MF)}

\addplot[color=myorange,mark=square*, line width=0.75pt, mark size=0.7] plot coordinates {
    (16,0.671)
    (32,0.684)
    (64,0.692)
    (128,0.702)
    
};
\addlegendentry{Learned Similarity (MLP)}

    
\addplot[color=mygreen, line width=0.75pt,mark=triangle*,mark options={rotate=180}, mark size=0.7] plot coordinates {
    (24,0.687748344370861)
    (48,0.6905629139072847)
    (96,0.6991721854304636)
    (192,0.6951986754966888)
    
};
\addlegendentry{MLP+GMF (NeuMF)}

\addplot[color=myred,mark=triangle*, line width=0.75pt, mark size=0.7] plot coordinates {
    (24,0.684)
    (48,0.707)
    (96,0.726)
    (192,0.730)
    
};
\addlegendentry{MLP+GMF pretrained (NeuMF)}

\addlegendimage{only marks, mark=o, mark options={scale=1, fill=black}}
\addplot[only marks,mark=o, mark size=2, myblue] coordinates { 
(192,0.73096)
(128,0.730629)
(32,0.71821)
};
\addplot[only marks,mark=o, mark size=2, mygreen] coordinates { 
(192,0.6951986754966888)
(96,0.6991721854304636)
(48,0.6905629139072847)
(24,0.687748344370861)
};
\addlegendentry{Replicated experiment}

\end{axis}
\end{tikzpicture}
\qquad
\begin{tikzpicture}[scale=1,line width=1pt,every node/.style={outer sep=0}]
\begin{axis}[
    height=15em,
    width=0.47\textwidth,
    title={Movielens},
    legend style={nodes={scale=0.6, transform shape}},
    legend pos=south east,
    grid=major,
    xlabel=Embedding dimension,
    ylabel=NDCG@10,
    xmin=10, 
    xmax=256,
    xmode=log,
    log basis x={2},
    x tick label style={/pgf/number format/1000 sep=\,},
    log ticks with fixed point,
    ymin=0.30, ymax=0.46,  
    ytick={0.32,0.34,0.36,0.38,0.40,0.42,0.44,0.46},
    y tick label style={
        /pgf/number format/.cd,
        fixed,
        fixed zerofill,
        precision=3,
        /tikz/.cd
    }
            ]
\addplot[mark=*,myblue, line width=0.75pt, mark size=0.7] plot coordinates {
    (32,0.43783042787950005)
    (128,0.4507639884014693)
    (192,0.45447841992459775)
    
};
\addlegendentry{Matrix Factorization (MF)}

\addplot[color=myorange,mark=square*, line width=0.75pt, mark size=0.7] plot coordinates {
    (16,0.399)
    (32,0.410)
    (64,0.425)
    (128,0.426)
    
};
\addlegendentry{Learned Similarity (MLP)}

    
\addplot[color=mygreen, line width=0.75pt,mark=triangle*,mark options={rotate=180}, mark size=0.7] plot coordinates {
    (24,0.4091363807142061)
    (48,0.4163192760319735)
    (96,0.4178577268663804)
    (192,0.4210999174140757)
    
};
\addlegendentry{MLP+GMF (NeuMF)}

\addplot[color=myred,mark=triangle*, line width=0.75pt, mark size=0.7] plot coordinates {
    (24,0.403)
    (48,0.426)
    (96,0.445)
    (192,0.447)
    
};
\addlegendentry{MLP+GMF pretrained (NeuMF)}

\addlegendimage{only marks, mark=o, mark options={scale=1, fill=black}}
\addplot[only marks,mark=o, mark size=2, myblue] coordinates { 
(192,0.45447841992459775)
(128,0.4507639884014693)
(32,0.43783042787950005)
};
\addplot[only marks,mark=o, mark size=2, mygreen] coordinates { 
    (24,0.4091363807142061)
    (48,0.4163192760319735)
    (96,0.4178577268663804)
    (192,0.4210999174140757)
};
\addlegendentry{Replicated experiment}

\end{axis}
\end{tikzpicture}


\begin{tikzpicture}[scale=1,line width=1pt,every node/.style={outer sep=0}]
\begin{axis}[
    height=15em,
    width=0.47\textwidth,
    title={Pinterest},
    legend style={nodes={scale=0.6, transform shape}},
    legend pos=south east,
    grid=major,
    xlabel=Embedding dimension,
    ylabel=HR@10,
    xmin=10, 
    xmax=256,
    xmode=log,
    log basis x={2},
    x tick label style={/pgf/number format/1000 sep=\,},
    log ticks with fixed point,
    ymin=0.78, ymax=0.90,  
    ytick={0.78,0.80,0.82,0.84,0.86,0.88,0.90},
    y tick label style={
        /pgf/number format/.cd,
        fixed,
        fixed zerofill,
        precision=3,
        /tikz/.cd
    }
            ]
\addplot[mark=*,myblue, line width=0.75pt, mark size=0.7] plot coordinates {
    (32,0.886404406834943)
    (128,0.888741913856524)
    (192,0.889756645586823)
    
};
\addlegendentry{Matrix Factorization (MF)}

\addplot[color=myorange,mark=square*, line width=0.75pt, mark size=0.7] plot coordinates {
    (16,0.859)
    (32,0.865)
    (64,0.868)
    (128,0.869)
    
};
\addlegendentry{Learned Similarity (MLP)}

\addplot[color=mygreen, line width=0.75pt,mark=triangle*,mark options={rotate=180}, mark size=0.7] plot coordinates {
    (24,0.8684110388316089)
    (48,0.8703861416637977)
    (96,0.86859)
    (192,0.8614)
    
};
\addlegendentry{MLP+GMF (NeuMF)}

\addplot[color=myred,mark=triangle*, line width=0.75pt, mark size=0.7] plot coordinates {
    (24,0.878)
    (48,0.880)
    (96,0.879)
    (192,0.877)
    
};
\addlegendentry{MLP+GMF pretrained (NeuMF)}

\addlegendimage{only marks, mark=o, mark options={scale=1, fill=black}}
\addplot[only marks,mark=o, mark size=2, myblue] coordinates { 
(32,0.886404406834943)
(128,0.888741913856524)
(192,0.889756645586823)
};
\addplot[only marks,mark=o, mark size=2, mygreen] coordinates { 
(192,0.8614)
(96,0.86859)
(48,0.8703861416637977)
(24,0.8684110388316089)
};
\addlegendentry{Replicated experiment}

\end{axis}
\end{tikzpicture}
\qquad
\begin{tikzpicture}[scale=1,line width=1pt,every node/.style={outer sep=0}]
\begin{axis}[
    height=15em,
    width=0.47\textwidth,
    title={Pinterest},
    legend style={nodes={scale=0.6, transform shape}},
    legend pos=south east,
    grid=major,
    xlabel=Embedding dimension,
    ylabel=NDCG@10,
    xmin=10, 
    xmax=256,
    xmode=log,
    log basis x={2},
    x tick label style={/pgf/number format/1000 sep=\,},
    log ticks with fixed point,
    ymin=0.48, ymax=0.58,  
    ytick={0.48,0.50,0.52,0.54,0.56,0.58},
    y tick label style={
        /pgf/number format/.cd,
        fixed,
        fixed zerofill,
        precision=3,
        /tikz/.cd
    }
            ]
\addplot[mark=*,myblue, line width=0.75pt, mark size=0.7] plot coordinates {
    (32,0.5721797668903681)
    (128,0.57594)
    (192,0.57762)
    
};
\addlegendentry{Matrix Factorization (MF)}

\addplot[color=myorange,mark=square*, line width=0.75pt, mark size=0.7] plot coordinates {
    (16,0.536)
    (32,0.538)
    (64,0.542)
    (128,0.545)
    
};
\addlegendentry{Learned Similarity (MLP)}

\addplot[color=mygreen, line width=0.75pt,mark=triangle*,mark options={rotate=180}, mark size=0.7] plot coordinates {
    (24,0.5471230723019709)
    (48,0.5480010804854436)
    (96,0.5409494067702199)
    (192,0.5364656332694538)
    
};
\addlegendentry{MLP+GMF (NeuMF)}

\addplot[color=myred,mark=triangle*, line width=0.75pt, mark size=0.7] plot coordinates {
    (24,0.555)
    (48,0.558)
    (96,0.555)
    (192,0.552)
    
};
\addlegendentry{MLP+GMF pretrained (NeuMF)}

\addlegendimage{only marks, mark=o, mark options={scale=1, fill=black}}
\addplot[only marks,mark=o, mark size=2, myblue] coordinates { 
    (32,0.5721797668903681)
    (128,0.57594)
    (192,0.57762)
};
\addplot[only marks,mark=o, mark size=2, mygreen] coordinates { 
    (24,0.5471230723019709)
    (48,0.5480010804854436)
    (96,0.5409494067702199)
    (192,0.5364656332694538)
};
\addlegendentry{Replicated experiment}

\end{axis}
\end{tikzpicture}

\caption{Comparison of learned similarities (MLP, NeuMF) with a dot product: The results for MLP and pretrained NeuMF are from~\citet{rendle,ncf}. MF substantially outperforms MF, NeuMF, and pretrained NeuMF. Nonetheless, on \movielens, when considering large embeddings, pretrained NeuMF is competitive. }\label{fig:base_results}
\end{figure*}
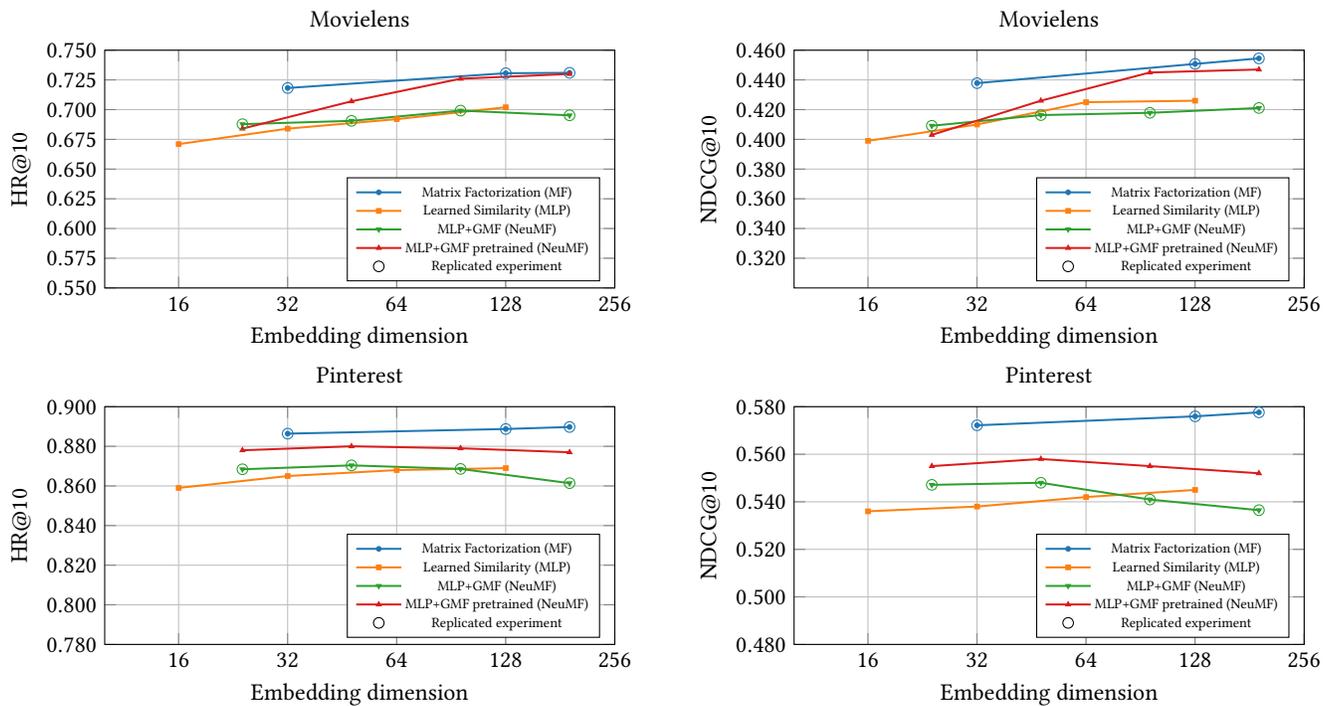  
Another finding (from~\citet{tois}) the careful reader can rediscover in our experiments is the \rp3b performance on the \pinterest dataset: although MF again demonstrates its higher accuracy, \rp3b demonstrates competitive performance overcoming all the remaining baselines. 
Overall, the general take-home message of~\citet{tois} experiments is confirmed: \textit{NeuMF is often not better than relatively simple and well-known techniques}.

Finally, Table~\ref{tab:vs_NeuMF} compares, for the sake of completeness, our experiments on NeuMF without pretraining with the same configuration from~\citet{ncf}. 
The results in columns marked with the reference are from Table 2 in~\citet{ncf} and correspond to the results for the NeuMF model without pretraining. 
As shown before qualitatively, the replicated results obtained through the benchmark framework overlap the original ones. 
However, considering $64$ factors on \pinterest, an appreciable difference can be observed that regards the nDCG value.
This is probably due to the non-deterministic initialization of the model that leads to slightly different results.
The effect seems to be more evident in the models with a greater embedding size, suggesting that the model accumulates the initial uncertainties. 
Remarkably, the deviation in the results exhibits a different trend (from the original model).
Even though this could be a signal of lack of robustness of the model, further investigation is needed to shed light on this behavior.

\begingroup
\setlength{\tabcolsep}{1pt}
\begin{table}
\centering
\caption{Performance of NeuMF without
pre-training. The table compares replicated experiments (on the left) with prior experiments~\citet{ncf}. Differently from~\citet{ncf}, the results on \pinterest show a performance decrease with 64 factors. All metrics are with cutoff @10.}
\label{tab:vs_NeuMF}
\small
\ra{1.4}
\resizebox{\linewidth}{!}{\begin{tabular}{l@{\hskip 6pt}cc@{\hskip 6pt}cc@{\hskip 6pt}cc@{\hskip 6pt}cc}\toprule
\multirow{2}{*}{Factors} & \multicolumn{2}{c@{\hskip 6pt}}{\movielens} & \multicolumn{2}{c@{\hskip 6pt}}{\movielens~\cite{ncf}}& \multicolumn{2}{c@{\hskip 6pt}}{\pinterest} & \multicolumn{2}{c}{\pinterest~\cite{ncf}}\\
\cmidrule(r{6pt}){2-3} \cmidrule(r{6pt}){4-5} \cmidrule(r{6pt}){6-7} \cmidrule{8-9}
& nDCG & HR & nDCG & HR& nDCG & HR & nDCG & HR\\ \midrule
{\phantom{0}}8~\cite{ncf} - {\phantom{0}}24~\cite{rendle} & 0.409& 0.688& 0.410 & 0.688 & 0.547 & 0.868& 0.546& 0.869\\
16~\cite{ncf} - {\phantom{0}}48~\cite{rendle} & 0.416& 0.691& 0.420 & 0.696& 0.548 &0.870 &0.547 & 0.871 \\
32~\cite{ncf} - {\phantom{0}}96~\cite{rendle} & 0.418& 0.699 & 0.425& 0.701& 0.541 & 0.869 & 0.549& 0.870\\
64~\cite{ncf} - 192~\cite{rendle} & 0.421 & 0.695 & 0.426& 0.705 & 0.536 & 0.861& 0.551& 0.872\\
\bottomrule
\end{tabular}}
\end{table}

\section{Comparing ANNs and MF on new contexts}


From now on, our investigation extends the previously described replicated experiments. 
These new experiments share the same setup of the previous section and exploit the same benchmark framework. 
The purpose is to provide a broader view of the experiments considering other evaluation dimensions.
First, we extend the list of metrics used to measure the accuracy of generated recommendation lists.
Second, we investigate beyond-accuracy evaluation dimensions, covering the novelty and diversity of the recommendations and the bias induced by the recommendation algorithms.
All the considered metrics have been implemented in Elliot\footnote{\url{https://github.com/sisinflab/elliot}}~\cite{DBLP:journals/corr/abs-2103-02590} and are publicly available\footnote{\url{https://github.com/sisinflab/Reenvisioning-the-comparison-between-Neural-Collaborative-Filtering-and-Matrix-Factorization}}. The specific nDCG formulation used in this paper is named \textit{nDCGRendle2020} to avoid confusion with the alternative implementation.

\subsection{An extended Accuracy evaluation}
The existence of a high correlation between the accuracy metrics has been recently shown~\cite{DBLP:journals/ir/ValcarceBPC20}.
Nevertheless, an evaluation that examines only HR and nDCG could be quite limited.
Therefore, we further extend the previous analysis considering other five metrics: F1-measure (F1)~\cite{DBLP:reference/sp/GunawardanaS15}, Limited Area Under the Curve (LAUC)~\cite{schroder2011setting}, Mean Average Precision (MAP)~\cite{DBLP:reference/sp/GunawardanaS15}, Mean Average Recall (MAR)~\cite{DBLP:reference/sp/GunawardanaS15}, and Mean Reciprocal Rank (MRR)~\cite{DBLP:conf/trec/Voorhees99}.

Table~\ref{tab:result_extra_acc} reports the results for the extended accuracy evaluation.
Observing the big picture, MF is still one of the most competitive models, consistently being the best model regarding all the considered metrics on \pinterest.
However, the situation is quite different on \movielens, since \easer shows the best performance in terms of MAP, MAR, and MRR.
Another interesting confirmation is the \rp3b performance on \pinterest. For all the considered metrics, it shows to be the second-best model.
However, if we observe the outcomes on \movielens, the situation is much more confusing.
Previous experiments showed that MF was the most accurate method, followed by \easer.
Table~\ref{tab:result_extra_acc} shows a quite different scenario, with \easer being the best model regarding MAP, MAR, and MRR, and the second best concerning the remaining metrics.
Conversely, MF still shows competitive results, but regarding MAP and MRR, it is not in the first two places.
Overall, MF, Slim, and iALS outperform NeuMF on these two datasets, hence confirming the most important finding of the previous experiment.

\subsubsection{Statistical hypothesis tests}
To complete the study regarding the accuracy evaluation, we investigated whether the differences between the accuracy results of the various methods are statistically significant.
Figure~\ref{fig:stattest} shows eight heatmaps of statistical significance calculated with the Student's paired t-test.
Statistically significant differences (with a p-value lower than 0.05) are drawn in green.
In contrast, p-values greater than or equal to the 0.05 threshold value are colored by shades of red.

\begin{figure*}
\centering
\begin{tabular}{cccc} 
  \includegraphics[width=38mm]{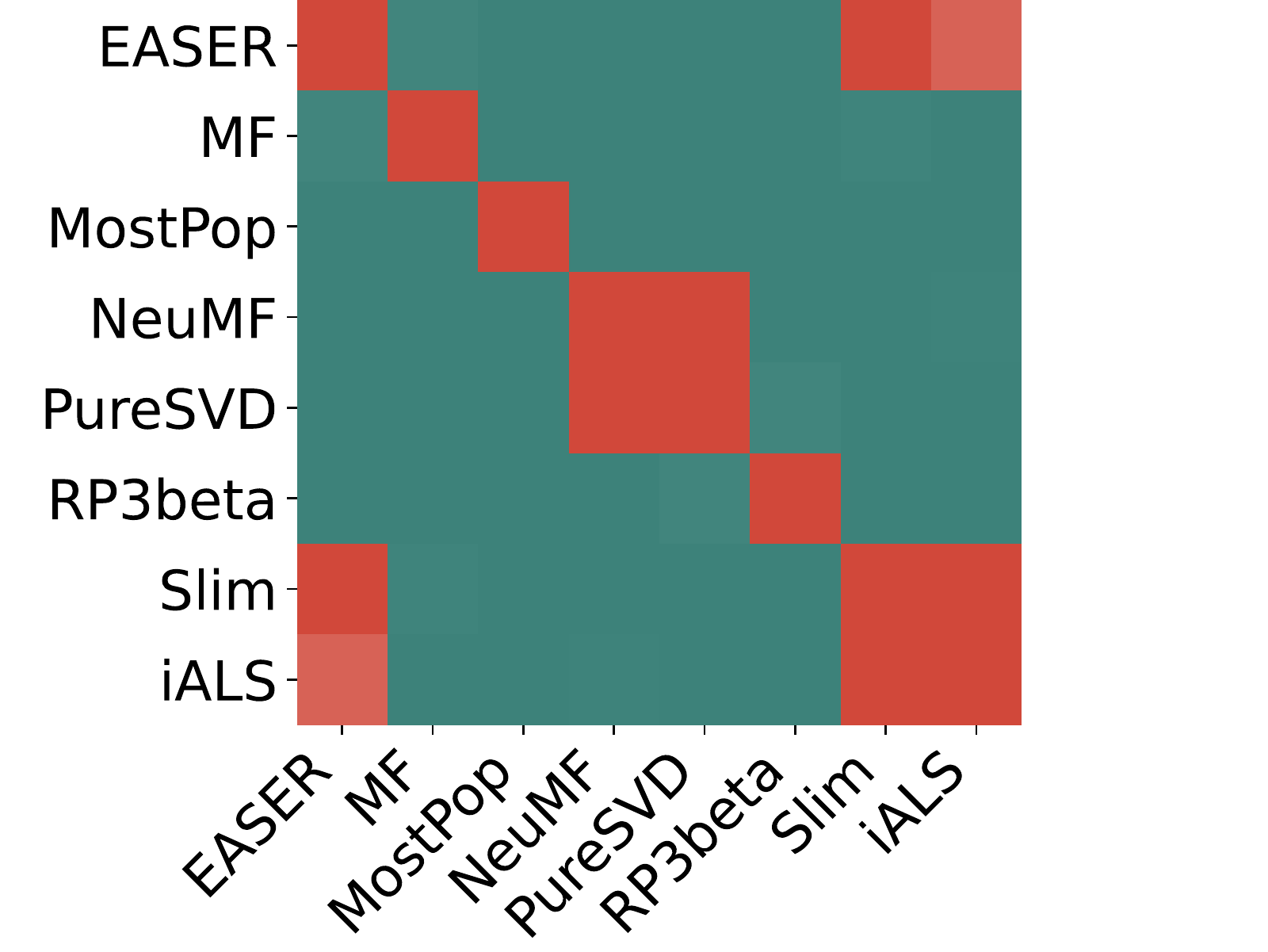} &   
  \includegraphics[width=38mm]{figures/ml1m_paired_HR.pdf} &  
  \includegraphics[width=38mm]{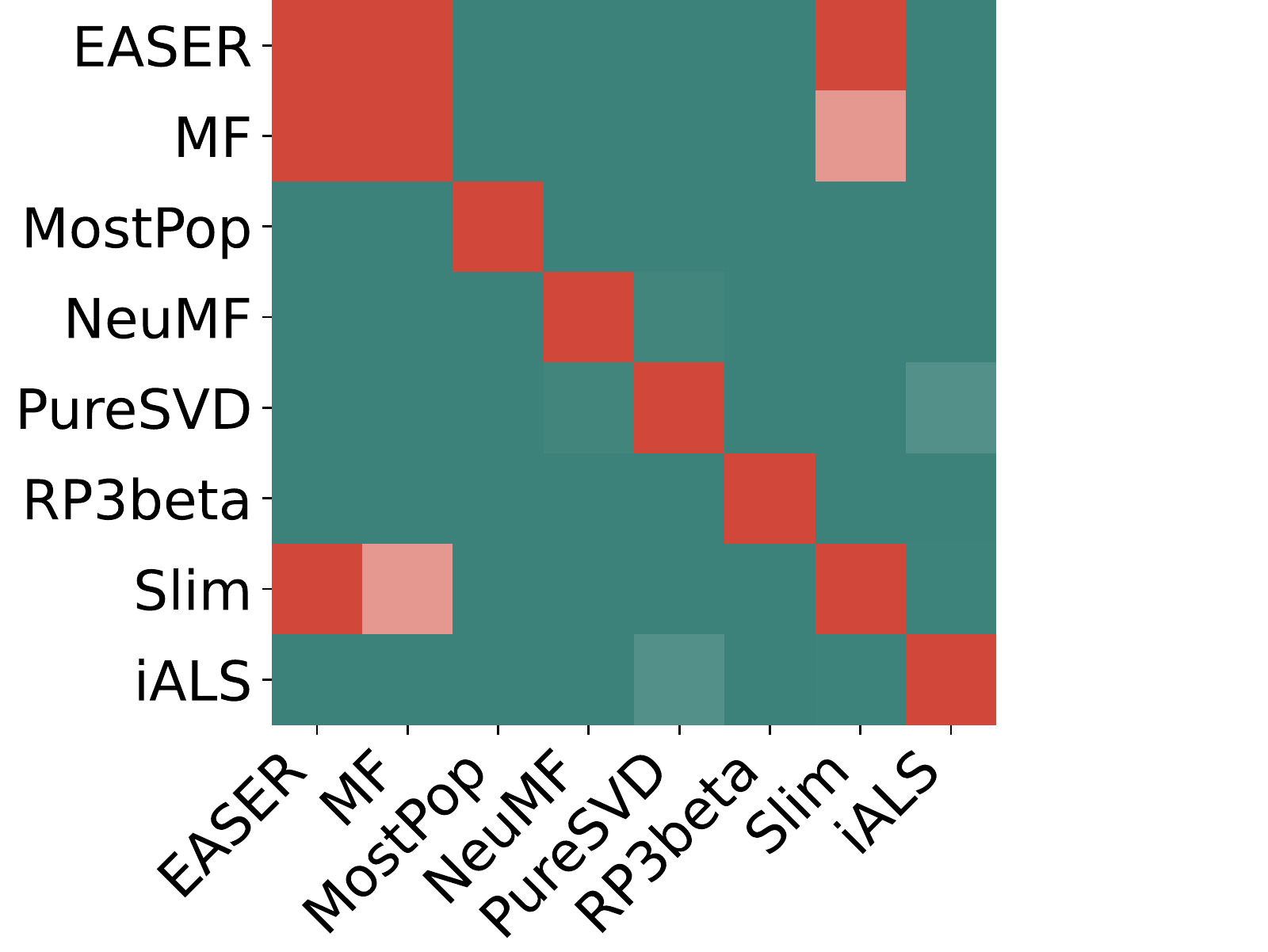} &   
  \includegraphics[width=38mm]{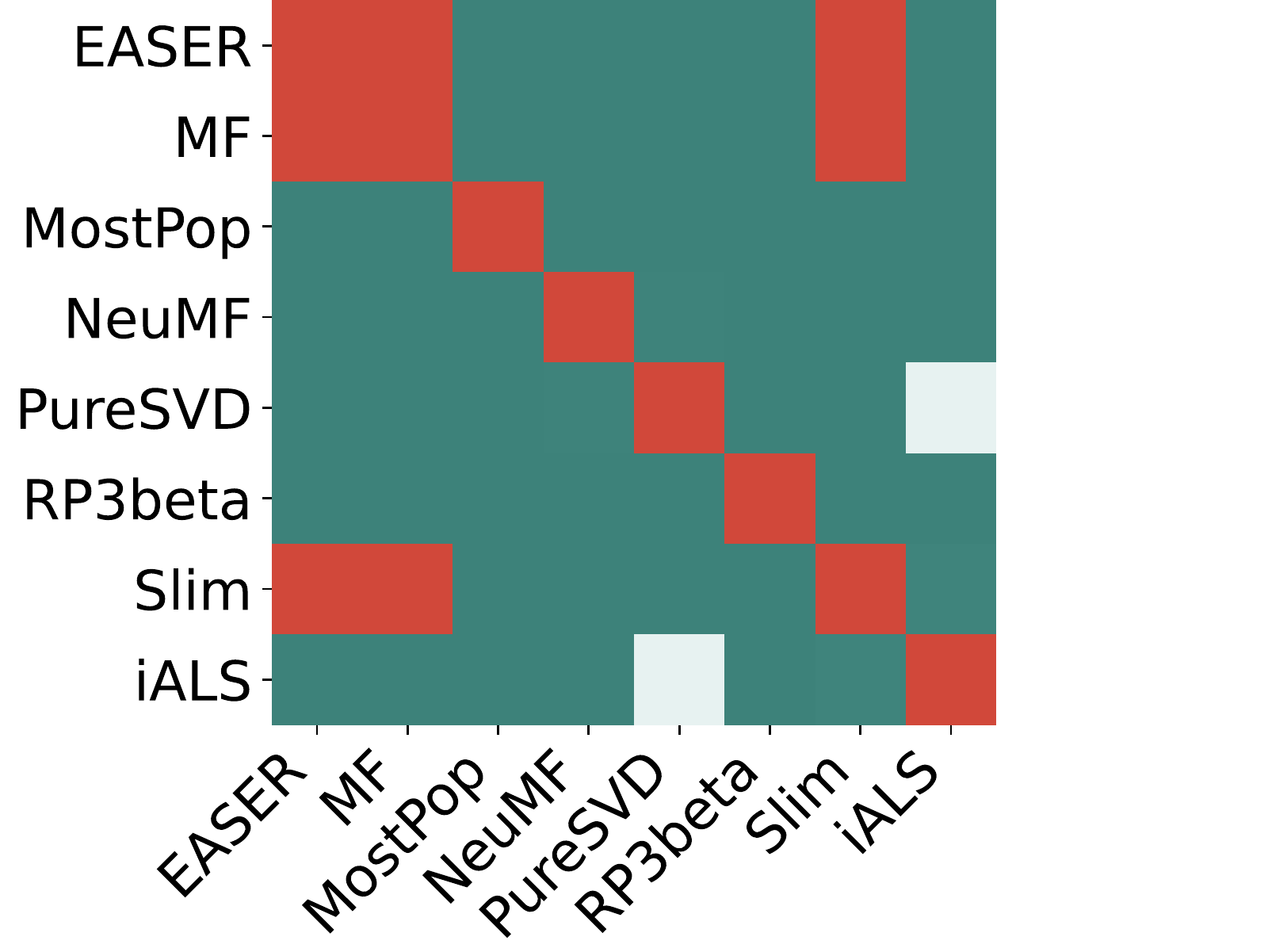}
 \\
nDCG & HR & MAP& MRR\\
\multicolumn{4}{c}{\movielens}\\[6pt] 
  \includegraphics[width=38mm]{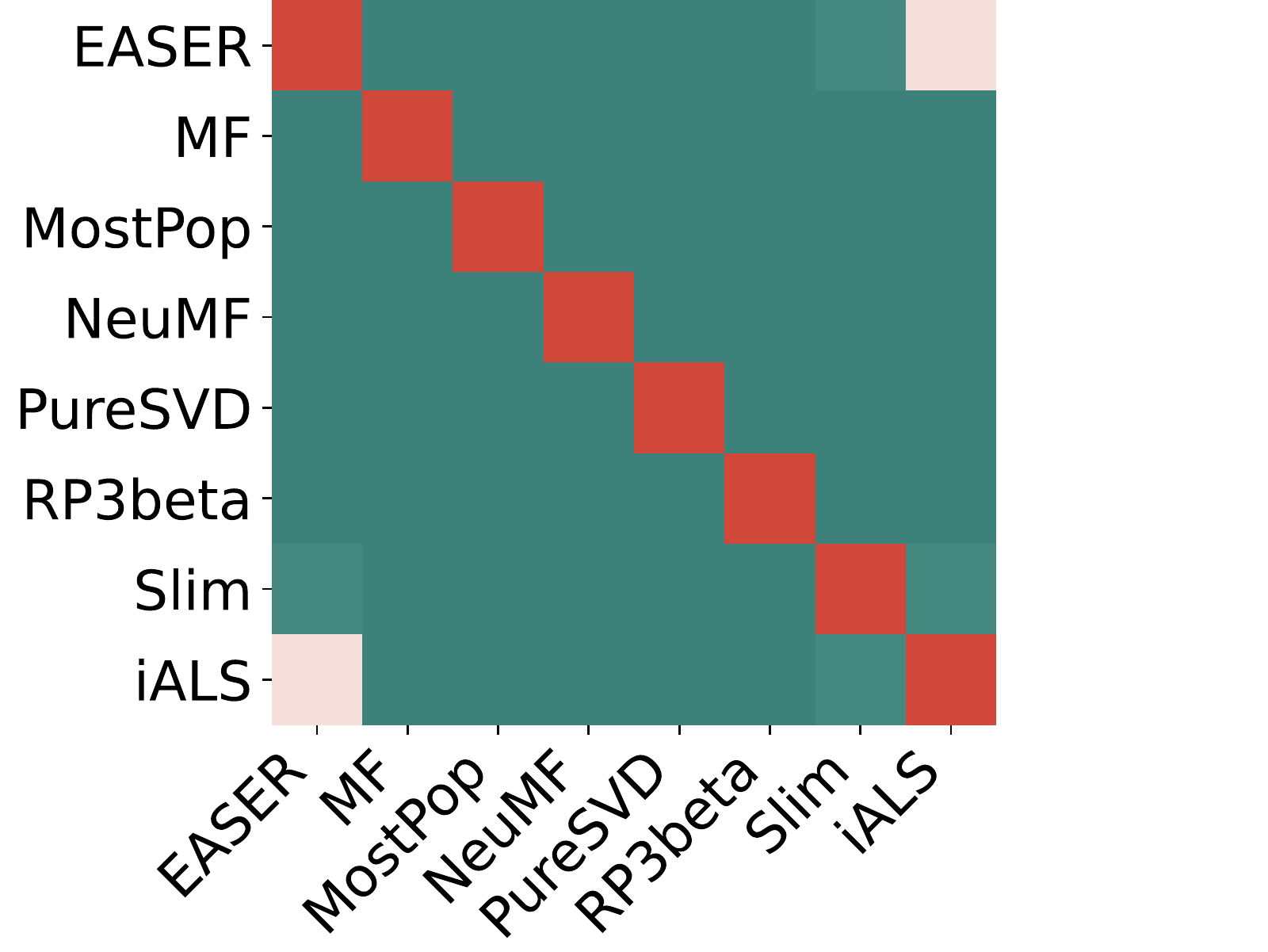} &   
  \includegraphics[width=38mm]{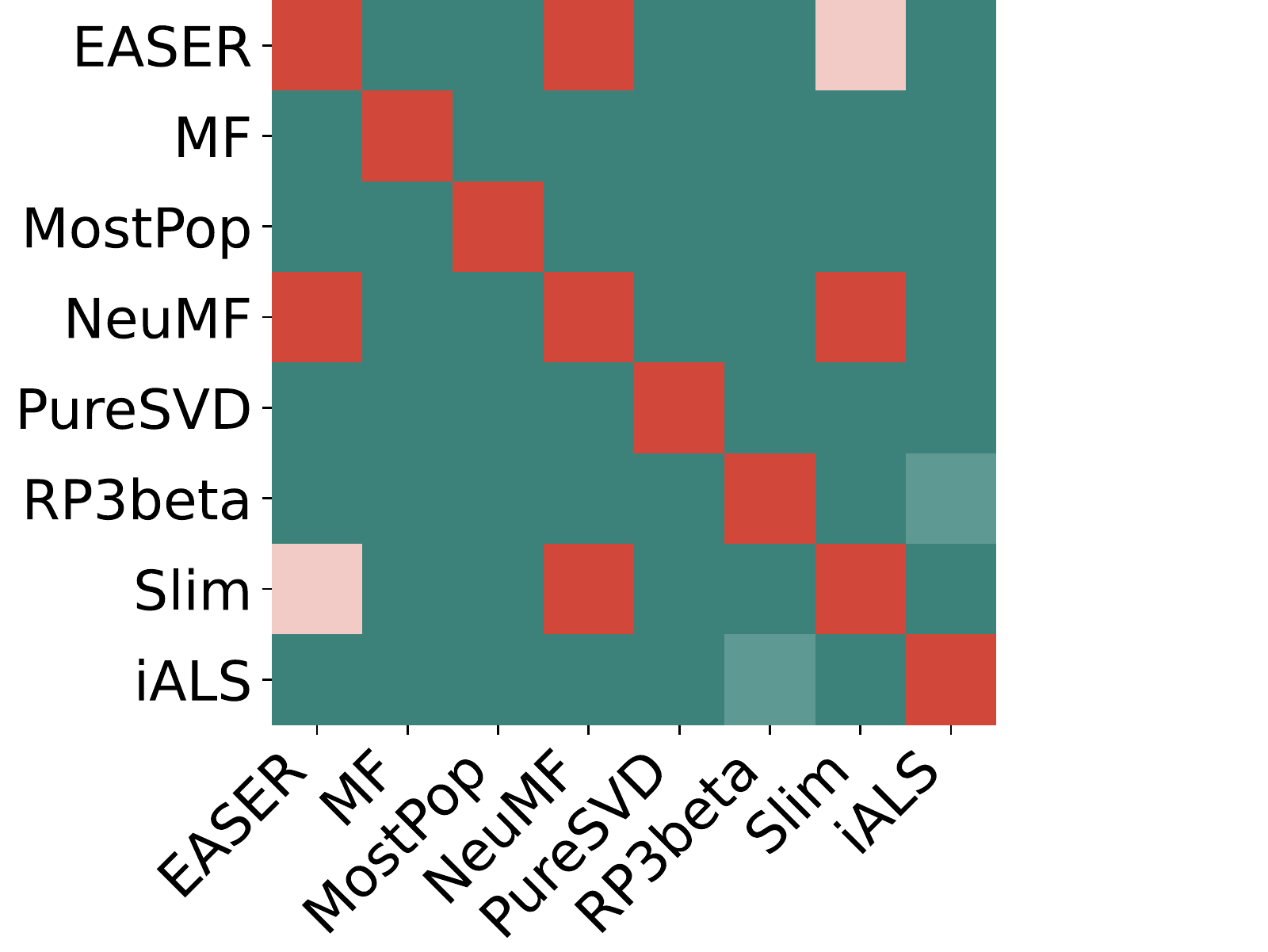} &  
  \includegraphics[width=38mm]{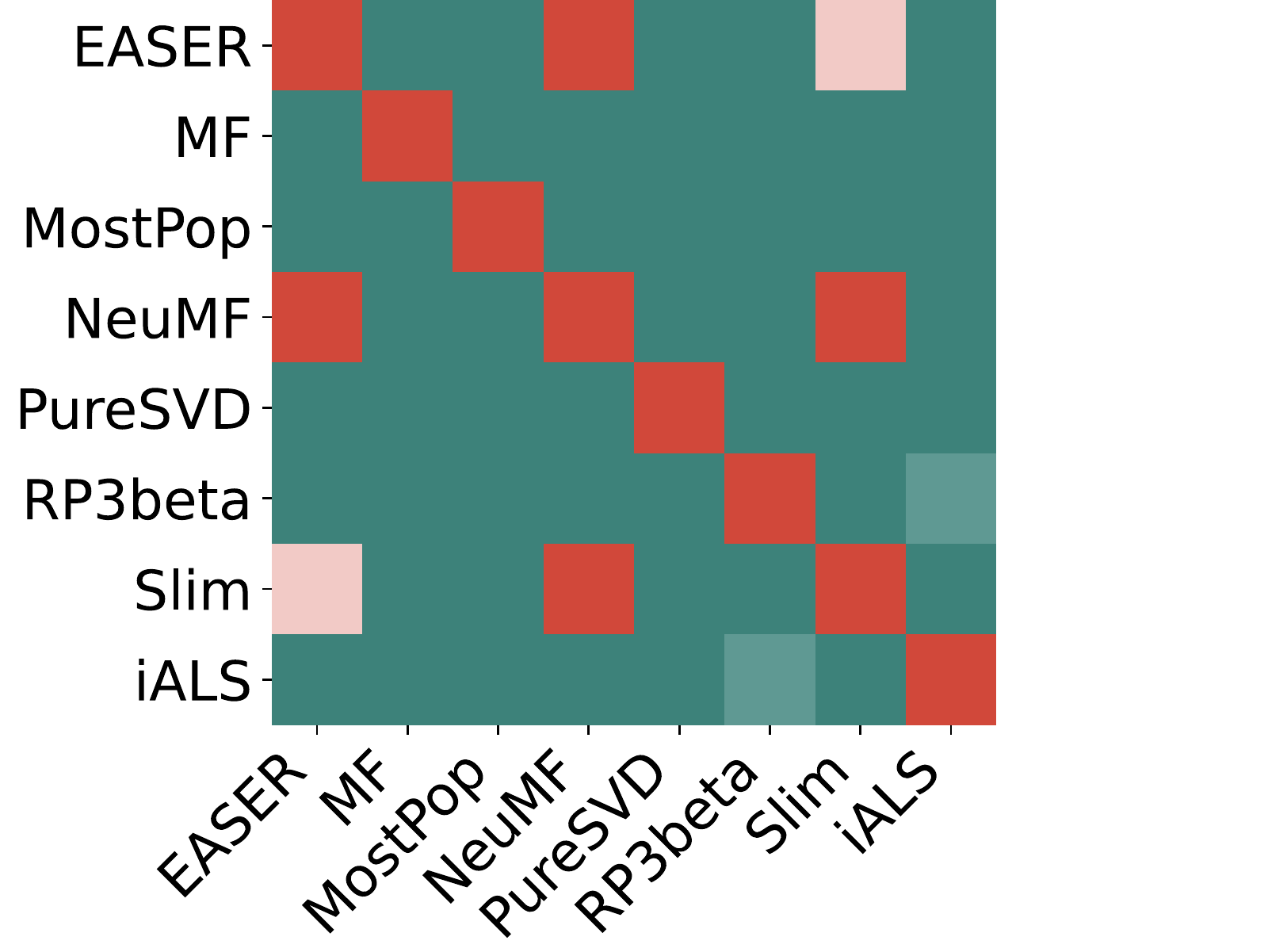} &
  \includegraphics[width=38mm]{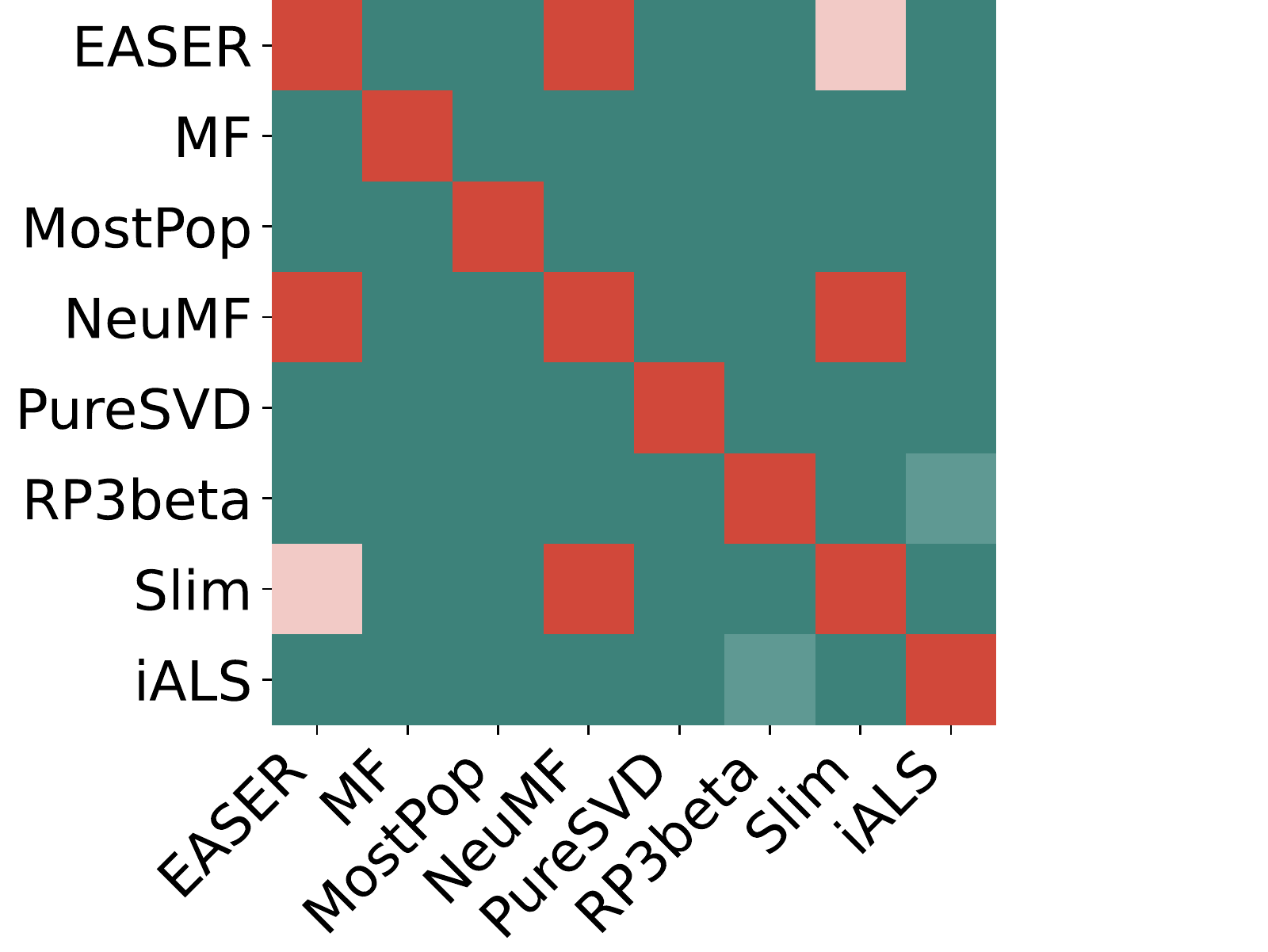} \\
nDCG & HR& F1& LAUC\\
\multicolumn{4}{c}{\pinterest}
\end{tabular}
\caption{Statistical hypothesis tests using Student's paired t-test with a threshold value (light red) of p=0.05. Algorithm pairs which results are statistically significant are in green, the results that are not statistically significant are in red.}\label{fig:stattest}
\end{figure*}
Figure~\ref{fig:stattest} confirms that MF significantly overcomes the other methods regarding nDCG and HR on both \movielens and \pinterest.
Besides MF, and considering the same metrics, the differences between \easer, iALS, and Slim are not always statistically significant.
Moreover, when analyzing MAP and MRR, it is noteworthy that the difference between \easer, Slim, and MF are not significant.
For what concerns NeuMF, the situation is different. Indeed, the differences with \easer, PureSVD, and Slim are not always significant.
Finally, \rp3b deserves a concluding remark since all the differences with the other models are statistically significant, thus confirming its positive performance on \pinterest and the below-the-average one on \movielens.

\begingroup
\setlength{\tabcolsep}{1pt}
\begin{table}[h]
\centering
\caption{Comparison of NeuMF and MF with various baselines on an extended set of accuracy metrics with cutoff @10. The best results are highlighted in bold, the
second-best result is underlined.}\label{tab:result_extra_acc}
\small
\ra{1.4}
\resizebox{\linewidth}{!}{\begin{tabular}{l@{\hskip 6pt}ccccc@{\hskip 6pt}ccccc}\toprule
\multirow{2}{*}{Method} & \multicolumn{5}{c@{\hskip 20pt}}{\movielens}& \multicolumn{5}{c@{\hskip 20pt}}{\pinterest}\\
\cmidrule(l{2pt}r{4pt}){2-6} \cmidrule{7-11}
& F1 & LAUC & MAP & MAR & MRR& F1 & LAUC & MAP & MAR & MRR\\ \midrule
MostPop & 0.0825 & 0.4531 & 0.0647 & 0.3072 & 0.1937 
& 0.0499 & 0.2742 & 0.0341 & 0.1717 & 0.1009\\
SLIM & 0.1303& 0.7159 & \underline{0.1204}& 0.5372 & \underline{0.3648}& 
0.1581& 0.8694 & 0.1535& 0.6757& 0.4649\\
iALS & 0.1295& 0.7117 & 0.1172& 0.5288 & 0.3537& 
0.1594& 0.8764 & 0.1525& 0.6786& 0.4587\\
NeuMF & 0.1264& 0.6947 & 0.1120& 0.5106 & 0.3363& 
0.1583& 0.8702& 0.1487& 0.6653& 0.4472\\
MF & \textbf{0.1316}& \textbf{0.7232} & 0.1188& \underline{0.5383} & 0.3573& 
\textbf{0.1618}& \textbf{0.8896}& \textbf{0.1584}& \textbf{0.6958}& \textbf{0.4796}\\
\midrule
\easer & \underline{0.1308}& \underline{0.7187} & \textbf{0.1210}& \textbf{0.54202} & \textbf{0.3655}& 
0.1579& 0.8682& 0.1532& 0.6752& 0.4639\\
\rp3b & 0.1229& 0.6753 & 0.1053& 0.4853 & 0.3166& 
\underline{0.1599}& \underline{0.8794}& \underline{0.1554}& \underline{0.6836}& \underline{0.4710}\\
PureSVD & 0.1259& 0.6921 & 0.1153& 0.5178 & 0.3486& 
0.1502& 0.8259& 0.1422& 0.6339& 0.4286\\
\bottomrule
\end{tabular}}
\end{table}
\subsection{Novelty and Diversity}
Once it is established how accurate the various methods are, our study expands beyond the accuracy evaluation. 
This section focuses on the ability of the recommendation algorithms to propose unknown items (Novelty), on overall item coverage, and on the ability to suggest highly diversified recommendation lists.
For what concerns Novelty, we measure Expected Free Discovery (EFD)~\cite{DBLP:conf/recsys/VargasC11} and Expected Popularity Complement (EPC)~\cite{DBLP:conf/recsys/VargasC11}, which measure the ability of a recommendation system to recommend items from the long tail. 
Concerning aggregate diversity metrics, we adopt Item Coverage~\cite{DBLP:reference/sp/GunawardanaS15} that measures the overall number of items the recommender suggests to the population.
Finally, to measure how diversified the recommendation lists are, we exploit two popular distributional inequality metrics, the Gini Index (Gini)~\cite{DBLP:reference/sp/GunawardanaS15} and Shannon Entropy (SE)~\cite{DBLP:reference/sp/GunawardanaS15}.
The Gini Index is defined as 1 - Gini Index from~\citet{DBLP:reference/sp/GunawardanaS15}, so that a higher value corresponds to a greater degree of diversification.


Figure~\ref{fig:novdiv} shows twelve bar charts that compare MF and NeuMF with the other baselines regarding the six observed metrics on the two datasets.
Let the analysis focus on Novelty.
It is worth noticing that, even here, MF outperforms NeuMF and the other baselines since it generates recommendation lists with a larger number of items belonging to the long tail.
Conversely, NeuMF shows poor performance, and only \rp3b and Most Popular behave worse. 
In general, also other matrix factorization models such as iALS and SLIM are shown to be competitive against the other baselines under analysis.
However, under the perspective of recommendation Diversity, the scenario dramatically changes.
In fact, regarding Item Coverage, NeuMF is the best performing model on \movielens, and a very competitive one on \pinterest (the best one is \rp3b), suggesting a higher overall number of items present in the catalog.
Conversely, MF (and the other MF-based models) are not able to win the comparison.
For what regards recommendation list diversification, the Gini bar chart reveal a more clear ranking of the methods. 
Again, MF fails to be effective in terms of diversity, and, on \movielens, only \easer and \rp3b show lower results.
NeuMF shines neither on \movielens dataset nor on \pinterest dataset. However, in both cases, it shows a greater propensity to generate personalized lists than MF. 
Interestingly, on both datasets, the iALS model is particularly competitive regarding the two distributional inequality metrics.
Finally, even here, the reader may appreciate how different the \rp3b performance is on the two datasets.

\input{figures/nov_div}


\subsection{Analysis of Recommendation Biases}
In the final part of the study, we focus on how the recommendation algorithms induce or amplify bias into the recommendation lists.
Indeed, user-item interactions are often distributed unevenly over different groups of users and categories of items.
This could be due to various reasons ranging from the naturally varying user preferences to the existence of a recommendation system in the preference collection system. 
Recommendation algorithms can inherit or even amplify this imbalanced distribution, leading to various kinds of bias.
To examine the bias effect 
we consider five different metrics: Average Coverage of Long Tail items (ACLT)~\cite{DBLP:conf/flairs/AbdollahpouriBM19}, Average Percentage of Long Tail Items (APLT)~\cite{DBLP:conf/flairs/AbdollahpouriBM19, DBLP:conf/recsys/AbdollahpouriBM17}, Average Recommendation Popularity (ARP)~\cite{DBLP:conf/flairs/AbdollahpouriBM19, DBLP:journals/pvldb/YinCLYC12}, Ranking-based Statistical Parity (RSP)~\cite{DBLP:conf/sigir/ZhuWC20}, and Ranking-based
Equal-Opportunity (REO)~\cite{DBLP:conf/sigir/ZhuWC20}.


Figure~\ref{fig:bias} shows ten bar charts that compare MF and NeuMF against the other baselines regarding these five bias measures on the two datasets.
The most straightforward metric to analyze is ARP.
This metric measures the average popularity of the recommended items in each list.
Interestingly, MF and NeuMF behave similarly on \movielens, while \easer and \rp3b are more prone to suggest popular items.
In contrast, the other MF-based methods, iALS, Slim, and PureSVD, show the best performance.
However, on \pinterest, the ranking is less clear since all the methods behave in a similar way.
Let the analysis focus on ACLT and APLT.
APLT measures the average percentage of long-tail items in the recommended lists, while ACLT measures how much exposure long-tail items get in the recommendations.
These two metrics exhibit three interesting behaviors: (i) both iALS and NeuMF seem to be less prone to these kinds of biases, (ii) MF, \easer, and PureSVD show to be heavily affected by them, (iii) the difference of \rp3b performance on the datasets influences the bias of the generated recommendations.

Finally, we focus our investigation on RSP and REO.
RSP measures whether items in different groups have the same probabilities of being recommended. Poor RSP means one or more groups have lower recommendation probabilities than others.
REO measures the bias that items in one or more groups have lower recommendation probabilities given the items enjoyed by users.
Differently from RSP, REO-based bias does not depend on sensitive attributes.

In this study, even though additional information could be retrieved to form item groups, the purpose is to conduct the investigation based on the same information available to the original authors.
Therefore we formed two distinct groups of items based on the popularity signal.
One group comprises the 20\% most popular items, while the other includes the remaining items.
For this reason, in the following, we refer to them as PopRSP and PopREO.

On \movielens dataset, iALS and SLIM exhibit the best performance regarding both metrics.
Even here, NeuMF demonstrates to be less prone than MF to this type of bias. 
MF does not show unsatisfactory results regarding both statistical parity and equal opportunity, but it never overcomes NeuMF.
Finally, \easer, \rp3b, and PureSVD are affected by the bias and under-recommended items from minority groups, even though these items are present in the user history.
In contrast, on \pinterest, \rp3b shows leading performance, along with iALS and NeuMF.
As detailed in Section~\ref{sec:settings}, following~\citet{ncf} and~\citet{rendle}, all the recommenders were optimized for accuracy. It is left as future work an extended analysis where the effect of different optimization goals could have on the recommendation accuracy and beyond-accuracy dimensions, as in~\citet{DBLP:journals/tiis/KaminskasB17}.
\input{figures/bias}


\section{Conclusion}

Understanding how the different recommendation algorithms work under unique evaluation dimensions is critical to advance the field.
In this work, we aimed to shed some light on this aspect, by contrasting recent models that are competitive against Neural Network approaches under complementary dimensions --- not only accuracy, but novelty, diversity, coverage, and bias.
In particular, we focus on the methods presented in~\citet{rendle,ncf}, and complemented our experimental exploration with the extensive analysis done in~\citet{tois}.
We have been able to replicate most of the results reported in those papers, where NeuMF is outperformed by the MF variation presented in \citet{rendle}.
Moreover, when reproducing these approaches in new contexts, such as other evaluation dimensions or more accuracy metrics, baselines like \easer and \rp3b are confirmed as solid candidates to be included in any comparison in the future, as their performance in terms of accuracy, diversity, and novelty is sometimes better than those of neural network approaches.
However, it is important to highlight that the trend obtained for NeuMF is slightly different than from the original paper and, in particular, our extended analysis on accuracy evidences that the difference between this method and other baselines are not always statistically significant.

Our experiments have summarized and re-evaluated results from 3 recent papers, but they can be complemented in several ways.
For example, one direction that has been unexplored so far is the effect that the splitting methodology or the item selection strategy could have in all these methods. 
Recent research has evidenced that how items are selected may affect the evaluation results~\cite{DBLP:conf/recsys/CanamaresC20}; however, because we wanted to replicate the exact conditions of these papers, we did not change these experimental settings. It will be interesting to analyze this aspect and how it (may) change the ranking of the methods.
Another potential venue to improve this comparison is on the selection of datasets. Again, as we wanted to replicate the original papers, we were limited to use \movielens and \pinterest, however, it is crucial to understand how these methods work in other domains and under a wide array of evaluation dimensions, such as those explored here.

\begin{acks}
The authors acknowledge partial support of the projects: PON ARS01\_00876 BIO-D
Casa delle Tecnologie Emergenti della Città di Matera, PON ARS01\_00821 FLET4.0,
PIA Servizi Locali 2.0 H2020 Passapartout - Grant n. 101016956, PIA ERP4.0,
and PID2019-108965\\GB-I00, 
IPZS-PRJ4\_IA\_NORMATIVO.
\end{acks}

\bibliographystyle{ACM-Reference-Format}
\bibliography{bibliography}


\begin{thebibliography}{41}


\ifx \showCODEN    \undefined \def \showCODEN     #1{\unskip}     \fi
\ifx \showDOI      \undefined \def \showDOI       #1{#1}\fi
\ifx \showISBNx    \undefined \def \showISBNx     #1{\unskip}     \fi
\ifx \showISBNxiii \undefined \def \showISBNxiii  #1{\unskip}     \fi
\ifx \showISSN     \undefined \def \showISSN      #1{\unskip}     \fi
\ifx \showLCCN     \undefined \def \showLCCN      #1{\unskip}     \fi
\ifx \shownote     \undefined \def \shownote      #1{#1}          \fi
\ifx \showarticletitle \undefined \def \showarticletitle #1{#1}   \fi
\ifx \showURL      \undefined \def \showURL       {\relax}        \fi
\providecommand\bibfield[2]{#2}
\providecommand\bibinfo[2]{#2}
\providecommand\natexlab[1]{#1}
\providecommand\showeprint[2][]{arXiv:#2}

\bibitem[\protect\citeauthoryear{Abdollahpouri, Burke, and
  Mobasher}{Abdollahpouri et~al\mbox{.}}{2017}]%
        {DBLP:conf/recsys/AbdollahpouriBM17}
\bibfield{author}{\bibinfo{person}{Himan Abdollahpouri}, \bibinfo{person}{Robin
  Burke}, {and} \bibinfo{person}{Bamshad Mobasher}.}
  \bibinfo{year}{2017}\natexlab{}.
\newblock \showarticletitle{Controlling Popularity Bias in Learning-to-Rank
  Recommendation}. In \bibinfo{booktitle}{\emph{RecSys}}.
  \bibinfo{publisher}{{ACM}}, \bibinfo{pages}{42--46}.
\newblock


\bibitem[\protect\citeauthoryear{Abdollahpouri, Burke, and
  Mobasher}{Abdollahpouri et~al\mbox{.}}{2019}]%
        {DBLP:conf/flairs/AbdollahpouriBM19}
\bibfield{author}{\bibinfo{person}{Himan Abdollahpouri}, \bibinfo{person}{Robin
  Burke}, {and} \bibinfo{person}{Bamshad Mobasher}.}
  \bibinfo{year}{2019}\natexlab{}.
\newblock \showarticletitle{Managing Popularity Bias in Recommender Systems
  with Personalized Re-Ranking}. In \bibinfo{booktitle}{\emph{{FLAIRS}
  Conference}}. \bibinfo{publisher}{{AAAI} Press}, \bibinfo{pages}{413--418}.
\newblock


\bibitem[\protect\citeauthoryear{Anelli, Bellog{\'{\i}}n, Ferrara, Malitesta,
  Merra, Pomo, Donini, and Noia}{Anelli et~al\mbox{.}}{2021}]%
        {DBLP:journals/corr/abs-2103-02590}
\bibfield{author}{\bibinfo{person}{Vito~Walter Anelli},
  \bibinfo{person}{Alejandro Bellog{\'{\i}}n}, \bibinfo{person}{Antonio
  Ferrara}, \bibinfo{person}{Daniele Malitesta},
  \bibinfo{person}{Felice~Antonio Merra}, \bibinfo{person}{Claudio Pomo},
  \bibinfo{person}{Francesco~Maria Donini}, {and} \bibinfo{person}{Tommaso~Di
  Noia}.} \bibinfo{year}{2021}\natexlab{}.
\newblock \showarticletitle{Elliot: {A} Comprehensive and Rigorous Framework
  for Reproducible Recommender Systems Evaluation}. In
  \bibinfo{booktitle}{\emph{{SIGIR}}}. \bibinfo{publisher}{{ACM}},
  \bibinfo{pages}{2405--2414}.
\newblock


\bibitem[\protect\citeauthoryear{Anelli, Di~Noia, Di~Sciascio, Ragone, and
  Trotta}{Anelli et~al\mbox{.}}{2020}]%
        {AnelliMinorTKDE}
\bibfield{author}{\bibinfo{person}{Vito~Walter Anelli},
  \bibinfo{person}{Tommaso Di~Noia}, \bibinfo{person}{Eugenio Di~Sciascio},
  \bibinfo{person}{Azzurra Ragone}, {and} \bibinfo{person}{Joseph Trotta}.}
  \bibinfo{year}{2020}\natexlab{}.
\newblock \showarticletitle{Semantic Interpretation of Top-N Recommendations}.
\newblock \bibinfo{journal}{\emph{IEEE Transactions on Knowledge and Data
  Engineering}} (\bibinfo{year}{2020}).
\newblock


\bibitem[\protect\citeauthoryear{Anelli, Noia, Sciascio, Ragone, and
  Trotta}{Anelli et~al\mbox{.}}{2019}]%
        {DBLP:conf/semweb/AnelliNSRT19}
\bibfield{author}{\bibinfo{person}{Vito~Walter Anelli},
  \bibinfo{person}{Tommaso~Di Noia}, \bibinfo{person}{Eugenio~Di Sciascio},
  \bibinfo{person}{Azzurra Ragone}, {and} \bibinfo{person}{Joseph Trotta}.}
  \bibinfo{year}{2019}\natexlab{}.
\newblock \showarticletitle{How to Make Latent Factors Interpretable by Feeding
  Factorization Machines with Knowledge Graphs}. In
  \bibinfo{booktitle}{\emph{{ISWC} {(1)}}} \emph{(\bibinfo{series}{Lecture
  Notes in Computer Science}, Vol.~\bibinfo{volume}{11778})}.
  \bibinfo{publisher}{Springer}, \bibinfo{pages}{38--56}.
\newblock


\bibitem[\protect\citeauthoryear{Ca{\~{n}}amares and Castells}{Ca{\~{n}}amares
  and Castells}{2020}]%
        {DBLP:conf/recsys/CanamaresC20}
\bibfield{author}{\bibinfo{person}{Roc{\'{\i}}o Ca{\~{n}}amares} {and}
  \bibinfo{person}{Pablo Castells}.} \bibinfo{year}{2020}\natexlab{}.
\newblock \showarticletitle{On Target Item Sampling in Offline Recommender
  System Evaluation}. In \bibinfo{booktitle}{\emph{RecSys}}.
  \bibinfo{publisher}{{ACM}}, \bibinfo{pages}{259--268}.
\newblock


\bibitem[\protect\citeauthoryear{Cremonesi, Koren, and Turrin}{Cremonesi
  et~al\mbox{.}}{2010}]%
        {DBLP:conf/recsys/CremonesiKT10}
\bibfield{author}{\bibinfo{person}{Paolo Cremonesi}, \bibinfo{person}{Yehuda
  Koren}, {and} \bibinfo{person}{Roberto Turrin}.}
  \bibinfo{year}{2010}\natexlab{}.
\newblock \showarticletitle{Performance of recommender algorithms on top-n
  recommendation tasks}. In \bibinfo{booktitle}{\emph{RecSys}}.
  \bibinfo{publisher}{{ACM}}, \bibinfo{pages}{39--46}.
\newblock


\bibitem[\protect\citeauthoryear{Cybenko}{Cybenko}{1989}]%
        {DBLP:journals/mcss/Cybenko89}
\bibfield{author}{\bibinfo{person}{George Cybenko}.}
  \bibinfo{year}{1989}\natexlab{}.
\newblock \showarticletitle{Approximation by superpositions of a sigmoidal
  function}.
\newblock \bibinfo{journal}{\emph{Math. Control. Signals Syst.}}
  \bibinfo{volume}{2}, \bibinfo{number}{4} (\bibinfo{year}{1989}),
  \bibinfo{pages}{303--314}.
\newblock


\bibitem[\protect\citeauthoryear{Dacrema, Boglio, Cremonesi, and
  Jannach}{Dacrema et~al\mbox{.}}{2021}]%
        {tois}
\bibfield{author}{\bibinfo{person}{Maurizio~Ferrari Dacrema},
  \bibinfo{person}{Simone Boglio}, \bibinfo{person}{Paolo Cremonesi}, {and}
  \bibinfo{person}{Dietmar Jannach}.} \bibinfo{year}{2021}\natexlab{}.
\newblock \showarticletitle{A Troubling Analysis of Reproducibility and
  Progress in Recommender Systems Research}.
\newblock \bibinfo{journal}{\emph{{ACM} Trans. Inf. Syst.}}
  \bibinfo{volume}{39}, \bibinfo{number}{2} (\bibinfo{year}{2021}),
  \bibinfo{pages}{20:1--20:49}.
\newblock


\bibitem[\protect\citeauthoryear{Dacrema, Parroni, Cremonesi, and
  Jannach}{Dacrema et~al\mbox{.}}{2020}]%
        {DBLP:conf/cikm/DacremaPCJ20}
\bibfield{author}{\bibinfo{person}{Maurizio~Ferrari Dacrema},
  \bibinfo{person}{Federico Parroni}, \bibinfo{person}{Paolo Cremonesi}, {and}
  \bibinfo{person}{Dietmar Jannach}.} \bibinfo{year}{2020}\natexlab{}.
\newblock \showarticletitle{Critically Examining the Claimed Value of
  Convolutions over User-Item Embedding Maps for Recommender Systems}. In
  \bibinfo{booktitle}{\emph{{CIKM}}}. \bibinfo{publisher}{{ACM}},
  \bibinfo{pages}{355--363}.
\newblock


\bibitem[\protect\citeauthoryear{Fern{\'{a}}ndez{-}Tob{\'{\i}}as, Cantador,
  Tomeo, Anelli, and Noia}{Fern{\'{a}}ndez{-}Tob{\'{\i}}as
  et~al\mbox{.}}{2019}]%
        {DBLP:journals/umuai/Fernandez-Tobias19}
\bibfield{author}{\bibinfo{person}{Ignacio Fern{\'{a}}ndez{-}Tob{\'{\i}}as},
  \bibinfo{person}{Iv{\'{a}}n Cantador}, \bibinfo{person}{Paolo Tomeo},
  \bibinfo{person}{Vito~Walter Anelli}, {and} \bibinfo{person}{Tommaso~Di
  Noia}.} \bibinfo{year}{2019}\natexlab{}.
\newblock \showarticletitle{Addressing the user cold start with cross-domain
  collaborative filtering: exploiting item metadata in matrix factorization}.
\newblock \bibinfo{journal}{\emph{User Model. User Adapt. Interact.}}
  \bibinfo{volume}{29}, \bibinfo{number}{2} (\bibinfo{year}{2019}),
  \bibinfo{pages}{443--486}.
\newblock


\bibitem[\protect\citeauthoryear{Gunawardana and Shani}{Gunawardana and
  Shani}{2015}]%
        {DBLP:reference/sp/GunawardanaS15}
\bibfield{author}{\bibinfo{person}{Asela Gunawardana} {and}
  \bibinfo{person}{Guy Shani}.} \bibinfo{year}{2015}\natexlab{}.
\newblock \showarticletitle{Evaluating Recommender Systems}.
\newblock In \bibinfo{booktitle}{\emph{Recommender Systems Handbook}}.
  \bibinfo{publisher}{Springer}, \bibinfo{pages}{265--308}.
\newblock


\bibitem[\protect\citeauthoryear{He and Chua}{He and Chua}{2017}]%
        {DBLP:conf/sigir/0001C17}
\bibfield{author}{\bibinfo{person}{Xiangnan He} {and}
  \bibinfo{person}{Tat{-}Seng Chua}.} \bibinfo{year}{2017}\natexlab{}.
\newblock \showarticletitle{Neural Factorization Machines for Sparse Predictive
  Analytics}. In \bibinfo{booktitle}{\emph{{SIGIR}}}.
  \bibinfo{publisher}{{ACM}}, \bibinfo{pages}{355--364}.
\newblock


\bibitem[\protect\citeauthoryear{He, Liao, Zhang, Nie, Hu, and Chua}{He
  et~al\mbox{.}}{2017}]%
        {ncf}
\bibfield{author}{\bibinfo{person}{Xiangnan He}, \bibinfo{person}{Lizi Liao},
  \bibinfo{person}{Hanwang Zhang}, \bibinfo{person}{Liqiang Nie},
  \bibinfo{person}{Xia Hu}, {and} \bibinfo{person}{Tat{-}Seng Chua}.}
  \bibinfo{year}{2017}\natexlab{}.
\newblock \showarticletitle{Neural Collaborative Filtering}. In
  \bibinfo{booktitle}{\emph{{WWW}}}. \bibinfo{publisher}{{ACM}},
  \bibinfo{pages}{173--182}.
\newblock


\bibitem[\protect\citeauthoryear{Hu, Koren, and Volinsky}{Hu
  et~al\mbox{.}}{2008}]%
        {DBLP:conf/icdm/HuKV08}
\bibfield{author}{\bibinfo{person}{Yifan Hu}, \bibinfo{person}{Yehuda Koren},
  {and} \bibinfo{person}{Chris Volinsky}.} \bibinfo{year}{2008}\natexlab{}.
\newblock \showarticletitle{Collaborative Filtering for Implicit Feedback
  Datasets}. In \bibinfo{booktitle}{\emph{{ICDM}}}. \bibinfo{publisher}{{IEEE}
  Computer Society}, \bibinfo{pages}{263--272}.
\newblock


\bibitem[\protect\citeauthoryear{Jannach, de~Souza Pereira~Moreira, and
  Oldridge}{Jannach et~al\mbox{.}}{2020}]%
        {DBLP:conf/recsys/JannachMO20}
\bibfield{author}{\bibinfo{person}{Dietmar Jannach}, \bibinfo{person}{Gabriel
  de Souza Pereira~Moreira}, {and} \bibinfo{person}{Even Oldridge}.}
  \bibinfo{year}{2020}\natexlab{}.
\newblock \showarticletitle{Why Are Deep Learning Models Not Consistently
  Winning Recommender Systems Competitions Yet?: {A} Position Paper}. In
  \bibinfo{booktitle}{\emph{RecSys Challenge}}. \bibinfo{publisher}{{ACM}},
  \bibinfo{pages}{44--49}.
\newblock


\bibitem[\protect\citeauthoryear{Juan, Zhuang, Chin, and Lin}{Juan
  et~al\mbox{.}}{2016}]%
        {DBLP:conf/recsys/JuanZCL16}
\bibfield{author}{\bibinfo{person}{Yu{-}Chin Juan}, \bibinfo{person}{Yong
  Zhuang}, \bibinfo{person}{Wei{-}Sheng Chin}, {and}
  \bibinfo{person}{Chih{-}Jen Lin}.} \bibinfo{year}{2016}\natexlab{}.
\newblock \showarticletitle{Field-aware Factorization Machines for {CTR}
  Prediction}. In \bibinfo{booktitle}{\emph{RecSys}}.
  \bibinfo{publisher}{{ACM}}, \bibinfo{pages}{43--50}.
\newblock


\bibitem[\protect\citeauthoryear{Kaminskas and Bridge}{Kaminskas and
  Bridge}{2017}]%
        {DBLP:journals/tiis/KaminskasB17}
\bibfield{author}{\bibinfo{person}{Marius Kaminskas} {and}
  \bibinfo{person}{Derek Bridge}.} \bibinfo{year}{2017}\natexlab{}.
\newblock \showarticletitle{Diversity, Serendipity, Novelty, and Coverage: {A}
  Survey and Empirical Analysis of Beyond-Accuracy Objectives in Recommender
  Systems}.
\newblock \bibinfo{journal}{\emph{{ACM} Trans. Interact. Intell. Syst.}}
  \bibinfo{volume}{7}, \bibinfo{number}{1} (\bibinfo{year}{2017}),
  \bibinfo{pages}{2:1--2:42}.
\newblock


\bibitem[\protect\citeauthoryear{Koren}{Koren}{2008}]%
        {DBLP:conf/kdd/Koren08}
\bibfield{author}{\bibinfo{person}{Yehuda Koren}.}
  \bibinfo{year}{2008}\natexlab{}.
\newblock \showarticletitle{Factorization meets the neighborhood: a
  multifaceted collaborative filtering model}. In
  \bibinfo{booktitle}{\emph{{KDD}}}. \bibinfo{publisher}{{ACM}},
  \bibinfo{pages}{426--434}.
\newblock


\bibitem[\protect\citeauthoryear{Koren and Bell}{Koren and Bell}{2015}]%
        {DBLP:reference/sp/KorenB15}
\bibfield{author}{\bibinfo{person}{Yehuda Koren} {and}
  \bibinfo{person}{Robert~M. Bell}.} \bibinfo{year}{2015}\natexlab{}.
\newblock \showarticletitle{Advances in Collaborative Filtering}.
\newblock In \bibinfo{booktitle}{\emph{Recommender Systems Handbook}}.
  \bibinfo{publisher}{Springer}, \bibinfo{pages}{77--118}.
\newblock


\bibitem[\protect\citeauthoryear{Krichene and Rendle}{Krichene and
  Rendle}{2020}]%
        {DBLP:conf/kdd/KricheneR20}
\bibfield{author}{\bibinfo{person}{Walid Krichene} {and}
  \bibinfo{person}{Steffen Rendle}.} \bibinfo{year}{2020}\natexlab{}.
\newblock \showarticletitle{On Sampled Metrics for Item Recommendation}. In
  \bibinfo{booktitle}{\emph{{KDD}}}. \bibinfo{publisher}{{ACM}},
  \bibinfo{pages}{1748--1757}.
\newblock


\bibitem[\protect\citeauthoryear{Luo, Zhou, Xia, and Zhu}{Luo
  et~al\mbox{.}}{2014}]%
        {DBLP:journals/tii/LuoZXZ14}
\bibfield{author}{\bibinfo{person}{Xin Luo}, \bibinfo{person}{Mengchu Zhou},
  \bibinfo{person}{Yunni Xia}, {and} \bibinfo{person}{Qingsheng Zhu}.}
  \bibinfo{year}{2014}\natexlab{}.
\newblock \showarticletitle{An Efficient Non-Negative
  Matrix-Factorization-Based Approach to Collaborative Filtering for
  Recommender Systems}.
\newblock \bibinfo{journal}{\emph{{IEEE} Trans. Ind. Informatics}}
  \bibinfo{volume}{10}, \bibinfo{number}{2} (\bibinfo{year}{2014}),
  \bibinfo{pages}{1273--1284}.
\newblock


\bibitem[\protect\citeauthoryear{Ning and Karypis}{Ning and Karypis}{2011}]%
        {DBLP:conf/icdm/NingK11}
\bibfield{author}{\bibinfo{person}{Xia Ning} {and} \bibinfo{person}{George
  Karypis}.} \bibinfo{year}{2011}\natexlab{}.
\newblock \showarticletitle{{SLIM:} Sparse Linear Methods for Top-N Recommender
  Systems}. In \bibinfo{booktitle}{\emph{{ICDM}}}. \bibinfo{publisher}{{IEEE}
  Computer Society}, \bibinfo{pages}{497--506}.
\newblock


\bibitem[\protect\citeauthoryear{Paudel, Christoffel, Newell, and
  Bernstein}{Paudel et~al\mbox{.}}{2017}]%
        {DBLP:journals/tiis/PaudelCNB17}
\bibfield{author}{\bibinfo{person}{Bibek Paudel}, \bibinfo{person}{Fabian
  Christoffel}, \bibinfo{person}{Chris Newell}, {and} \bibinfo{person}{Abraham
  Bernstein}.} \bibinfo{year}{2017}\natexlab{}.
\newblock \showarticletitle{Updatable, Accurate, Diverse, and Scalable
  Recommendations for Interactive Applications}.
\newblock \bibinfo{journal}{\emph{{ACM} Trans. Interact. Intell. Syst.}}
  \bibinfo{volume}{7}, \bibinfo{number}{1} (\bibinfo{year}{2017}),
  \bibinfo{pages}{1:1--1:34}.
\newblock


\bibitem[\protect\citeauthoryear{Rendle}{Rendle}{2010}]%
        {DBLP:conf/icdm/Rendle10}
\bibfield{author}{\bibinfo{person}{Steffen Rendle}.}
  \bibinfo{year}{2010}\natexlab{}.
\newblock \showarticletitle{Factorization Machines}. In
  \bibinfo{booktitle}{\emph{{ICDM}}}. \bibinfo{publisher}{{IEEE} Computer
  Society}, \bibinfo{pages}{995--1000}.
\newblock


\bibitem[\protect\citeauthoryear{Rendle, Freudenthaler, Gantner, and
  Schmidt{-}Thieme}{Rendle et~al\mbox{.}}{2009}]%
        {DBLP:conf/uai/RendleFGS09}
\bibfield{author}{\bibinfo{person}{Steffen Rendle}, \bibinfo{person}{Christoph
  Freudenthaler}, \bibinfo{person}{Zeno Gantner}, {and} \bibinfo{person}{Lars
  Schmidt{-}Thieme}.} \bibinfo{year}{2009}\natexlab{}.
\newblock \showarticletitle{{BPR:} Bayesian Personalized Ranking from Implicit
  Feedback}. In \bibinfo{booktitle}{\emph{{UAI}}}. \bibinfo{publisher}{{AUAI}
  Press}, \bibinfo{pages}{452--461}.
\newblock


\bibitem[\protect\citeauthoryear{Rendle, Krichene, Zhang, and Anderson}{Rendle
  et~al\mbox{.}}{2020}]%
        {rendle}
\bibfield{author}{\bibinfo{person}{Steffen Rendle}, \bibinfo{person}{Walid
  Krichene}, \bibinfo{person}{Li Zhang}, {and} \bibinfo{person}{John~R.
  Anderson}.} \bibinfo{year}{2020}\natexlab{}.
\newblock \showarticletitle{Neural Collaborative Filtering vs. Matrix
  Factorization Revisited}. In \bibinfo{booktitle}{\emph{RecSys}}.
  \bibinfo{publisher}{{ACM}}, \bibinfo{pages}{240--248}.
\newblock


\bibitem[\protect\citeauthoryear{Said and Bellog{\'{\i}}n}{Said and
  Bellog{\'{\i}}n}{2014}]%
        {DBLP:conf/recsys/SaidB14}
\bibfield{author}{\bibinfo{person}{Alan Said} {and} \bibinfo{person}{Alejandro
  Bellog{\'{\i}}n}.} \bibinfo{year}{2014}\natexlab{}.
\newblock \showarticletitle{Comparative recommender system evaluation:
  benchmarking recommendation frameworks}. In
  \bibinfo{booktitle}{\emph{RecSys}}. \bibinfo{publisher}{{ACM}},
  \bibinfo{pages}{129--136}.
\newblock


\bibitem[\protect\citeauthoryear{Salakhutdinov and Mnih}{Salakhutdinov and
  Mnih}{2007}]%
        {DBLP:conf/nips/SalakhutdinovM07}
\bibfield{author}{\bibinfo{person}{Ruslan Salakhutdinov} {and}
  \bibinfo{person}{Andriy Mnih}.} \bibinfo{year}{2007}\natexlab{}.
\newblock \showarticletitle{Probabilistic Matrix Factorization}. In
  \bibinfo{booktitle}{\emph{{NIPS}}}. \bibinfo{publisher}{Curran Associates,
  Inc.}, \bibinfo{pages}{1257--1264}.
\newblock


\bibitem[\protect\citeauthoryear{Salakhutdinov and Mnih}{Salakhutdinov and
  Mnih}{2008}]%
        {DBLP:conf/icml/SalakhutdinovM08a}
\bibfield{author}{\bibinfo{person}{Ruslan Salakhutdinov} {and}
  \bibinfo{person}{Andriy Mnih}.} \bibinfo{year}{2008}\natexlab{}.
\newblock \showarticletitle{Bayesian probabilistic matrix factorization using
  Markov chain Monte Carlo}. In \bibinfo{booktitle}{\emph{{ICML}}}
  \emph{(\bibinfo{series}{{ACM} International Conference Proceeding Series},
  Vol.~\bibinfo{volume}{307})}. \bibinfo{publisher}{{ACM}},
  \bibinfo{pages}{880--887}.
\newblock


\bibitem[\protect\citeauthoryear{Schr{\"o}der, Thiele, and Lehner}{Schr{\"o}der
  et~al\mbox{.}}{2011}]%
        {schroder2011setting}
\bibfield{author}{\bibinfo{person}{Gunnar Schr{\"o}der}, \bibinfo{person}{Maik
  Thiele}, {and} \bibinfo{person}{Wolfgang Lehner}.}
  \bibinfo{year}{2011}\natexlab{}.
\newblock \showarticletitle{Setting Goals and Choosing Metrics for Recommender
  System Evaluations}. In \bibinfo{booktitle}{\emph{UCERSTI2 workshop at the
  5th ACM conference on recommender systems, Chicago, USA}},
  Vol.~\bibinfo{volume}{23}. \bibinfo{pages}{53}.
\newblock


\bibitem[\protect\citeauthoryear{Steck}{Steck}{2019}]%
        {DBLP:conf/www/Steck19}
\bibfield{author}{\bibinfo{person}{Harald Steck}.}
  \bibinfo{year}{2019}\natexlab{}.
\newblock \showarticletitle{Embarrassingly Shallow Autoencoders for Sparse
  Data}. In \bibinfo{booktitle}{\emph{{WWW}}}. \bibinfo{publisher}{{ACM}},
  \bibinfo{pages}{3251--3257}.
\newblock


\bibitem[\protect\citeauthoryear{Valcarce, Bellog{\'{\i}}n, Parapar, and
  Castells}{Valcarce et~al\mbox{.}}{2020}]%
        {DBLP:journals/ir/ValcarceBPC20}
\bibfield{author}{\bibinfo{person}{Daniel Valcarce}, \bibinfo{person}{Alejandro
  Bellog{\'{\i}}n}, \bibinfo{person}{Javier Parapar}, {and}
  \bibinfo{person}{Pablo Castells}.} \bibinfo{year}{2020}\natexlab{}.
\newblock \showarticletitle{Assessing ranking metrics in top-N recommendation}.
\newblock \bibinfo{journal}{\emph{Inf. Retr. J.}} \bibinfo{volume}{23},
  \bibinfo{number}{4} (\bibinfo{year}{2020}), \bibinfo{pages}{411--448}.
\newblock


\bibitem[\protect\citeauthoryear{Vargas and Castells}{Vargas and
  Castells}{2011}]%
        {DBLP:conf/recsys/VargasC11}
\bibfield{author}{\bibinfo{person}{Saul Vargas} {and} \bibinfo{person}{Pablo
  Castells}.} \bibinfo{year}{2011}\natexlab{}.
\newblock \showarticletitle{Rank and relevance in novelty and diversity metrics
  for recommender systems}. In \bibinfo{booktitle}{\emph{Proceedings of the
  2011 {ACM} Conference on Recommender Systems, RecSys 2011, Chicago, IL, USA,
  October 23-27, 2011}}, \bibfield{editor}{\bibinfo{person}{Bamshad Mobasher},
  \bibinfo{person}{Robin~D. Burke}, \bibinfo{person}{Dietmar Jannach}, {and}
  \bibinfo{person}{Gediminas Adomavicius}} (Eds.). \bibinfo{publisher}{{ACM}},
  \bibinfo{pages}{109--116}.
\newblock
\urldef\tempurl%
\url{https://dl.acm.org/citation.cfm?id=2043955}
\showURL{%
\tempurl}


\bibitem[\protect\citeauthoryear{Voorhees}{Voorhees}{1999}]%
        {DBLP:conf/trec/Voorhees99}
\bibfield{author}{\bibinfo{person}{Ellen~M. Voorhees}.}
  \bibinfo{year}{1999}\natexlab{}.
\newblock \showarticletitle{The {TREC-8} Question Answering Track Report}. In
  \bibinfo{booktitle}{\emph{{TREC}}} \emph{(\bibinfo{series}{{NIST} Special
  Publication}, Vol.~\bibinfo{volume}{500-246})}. \bibinfo{publisher}{National
  Institute of Standards and Technology {(NIST)}}.
\newblock


\bibitem[\protect\citeauthoryear{Xiao, Ye, He, Zhang, Wu, and Chua}{Xiao
  et~al\mbox{.}}{2017}]%
        {DBLP:conf/ijcai/XiaoY0ZWC17}
\bibfield{author}{\bibinfo{person}{Jun Xiao}, \bibinfo{person}{Hao Ye},
  \bibinfo{person}{Xiangnan He}, \bibinfo{person}{Hanwang Zhang},
  \bibinfo{person}{Fei Wu}, {and} \bibinfo{person}{Tat{-}Seng Chua}.}
  \bibinfo{year}{2017}\natexlab{}.
\newblock \showarticletitle{Attentional Factorization Machines: Learning the
  Weight of Feature Interactions via Attention Networks}. In
  \bibinfo{booktitle}{\emph{{IJCAI}}}. \bibinfo{publisher}{ijcai.org},
  \bibinfo{pages}{3119--3125}.
\newblock


\bibitem[\protect\citeauthoryear{Yin, Cui, Li, Yao, and Chen}{Yin
  et~al\mbox{.}}{2012}]%
        {DBLP:journals/pvldb/YinCLYC12}
\bibfield{author}{\bibinfo{person}{Hongzhi Yin}, \bibinfo{person}{Bin Cui},
  \bibinfo{person}{Jing Li}, \bibinfo{person}{Junjie Yao}, {and}
  \bibinfo{person}{Chen Chen}.} \bibinfo{year}{2012}\natexlab{}.
\newblock \showarticletitle{Challenging the Long Tail Recommendation}.
\newblock \bibinfo{journal}{\emph{Proc. {VLDB} Endow.}} \bibinfo{volume}{5},
  \bibinfo{number}{9} (\bibinfo{year}{2012}), \bibinfo{pages}{896--907}.
\newblock


\bibitem[\protect\citeauthoryear{Zhang, Yao, Sun, and Tay}{Zhang
  et~al\mbox{.}}{2019}]%
        {DBLP:journals/csur/ZhangYST19}
\bibfield{author}{\bibinfo{person}{Shuai Zhang}, \bibinfo{person}{Lina Yao},
  \bibinfo{person}{Aixin Sun}, {and} \bibinfo{person}{Yi Tay}.}
  \bibinfo{year}{2019}\natexlab{}.
\newblock \showarticletitle{Deep Learning Based Recommender System: {A} Survey
  and New Perspectives}.
\newblock \bibinfo{journal}{\emph{{ACM} Comput. Surv.}} \bibinfo{volume}{52},
  \bibinfo{number}{1} (\bibinfo{year}{2019}), \bibinfo{pages}{5:1--5:38}.
\newblock


\bibitem[\protect\citeauthoryear{Zhang, Lai, Zhang, Zhang, Liu, and Ma}{Zhang
  et~al\mbox{.}}{2014}]%
        {DBLP:conf/sigir/ZhangL0ZLM14}
\bibfield{author}{\bibinfo{person}{Yongfeng Zhang}, \bibinfo{person}{Guokun
  Lai}, \bibinfo{person}{Min Zhang}, \bibinfo{person}{Yi Zhang},
  \bibinfo{person}{Yiqun Liu}, {and} \bibinfo{person}{Shaoping Ma}.}
  \bibinfo{year}{2014}\natexlab{}.
\newblock \showarticletitle{Explicit factor models for explainable
  recommendation based on phrase-level sentiment analysis}. In
  \bibinfo{booktitle}{\emph{{SIGIR}}}. \bibinfo{publisher}{{ACM}},
  \bibinfo{pages}{83--92}.
\newblock


\bibitem[\protect\citeauthoryear{Zhu, Lin, He, Wang, Guan, Liu, and Cai}{Zhu
  et~al\mbox{.}}{2020a}]%
        {DBLP:journals/tkde/ZhuLHWGLC20}
\bibfield{author}{\bibinfo{person}{Yu Zhu}, \bibinfo{person}{Jinghao Lin},
  \bibinfo{person}{Shibi He}, \bibinfo{person}{Beidou Wang},
  \bibinfo{person}{Ziyu Guan}, \bibinfo{person}{Haifeng Liu}, {and}
  \bibinfo{person}{Deng Cai}.} \bibinfo{year}{2020}\natexlab{a}.
\newblock \showarticletitle{Addressing the Item Cold-Start Problem by
  Attribute-Driven Active Learning}.
\newblock \bibinfo{journal}{\emph{{IEEE} Trans. Knowl. Data Eng.}}
  \bibinfo{volume}{32}, \bibinfo{number}{4} (\bibinfo{year}{2020}),
  \bibinfo{pages}{631--644}.
\newblock


\bibitem[\protect\citeauthoryear{Zhu, Wang, and Caverlee}{Zhu
  et~al\mbox{.}}{2020b}]%
        {DBLP:conf/sigir/ZhuWC20}
\bibfield{author}{\bibinfo{person}{Ziwei Zhu}, \bibinfo{person}{Jianling Wang},
  {and} \bibinfo{person}{James Caverlee}.} \bibinfo{year}{2020}\natexlab{b}.
\newblock \showarticletitle{Measuring and Mitigating Item Under-Recommendation
  Bias in Personalized Ranking Systems}. In
  \bibinfo{booktitle}{\emph{{SIGIR}}}. \bibinfo{publisher}{{ACM}},
  \bibinfo{pages}{449--458}.
\newblock


\end{thebibliography}






\end{document}